\documentclass[pre,aps,reprint,superscriptaddress,floatflix,twocolumn,amsfonts,amsmath]{revtex4-1}
\usepackage{amssymb}
\usepackage{amsmath}
\usepackage{graphicx}
\usepackage{epsfig}

\newcommand{\stl}[1]{\mbox{$ \hspace{0.1em}
      \stackrel{\rule{0.4pt}{0.275ex}\hspace{0.40em} \!\!\!
      \overline{\hspace{0.06em}\vphantom{\rule{0.4pt}{0.0ex}}
      \hphantom{\mbox{$\displaystyle #1$}}
      \hspace{0.06em}  } \!\!\!\hspace{0.40em}\rule{0.4pt}{0.275ex}}
      {#1}\hspace{0.2em}$}}
\def\onedot{$\mathsurround0pt\ldotp$}
\def\cddot{
  \mathbin{\vcenter{\baselineskip.67ex
    \hbox{\onedot}\hbox{\onedot}}%
  }}%
\def\cdddot#1{
  \mathbin{\vcenter{\baselineskip.67ex
    \hbox{\onedot}\hbox{\onedot}\hbox{\onedot}%
  }}%
}

\begin{document}
\title{Controlling motile disclinations in a thick nematogenic material with an electric field}
\date{\today}
\def\iisc{\affiliation{Centre for Condensed Matter Theory, Department of Physics, Indian Institute
                       of Science, Bangalore 560064, India}}
\def\asu{\affiliation{Asutosh College, University of Calcutta, Kolkata 700026, India}}
\author{Amit Kumar Bhattacharjee}
\email{Email address: amitb@physics.iisc.ernet.in}\asu \iisc

\begin{abstract}
Manipulating topological disclination networks that arise in a symmetry-breaking phase transformation 
in widely varied systems including anisotropic materials can potentially lead to the design of novel 
materials like conductive microwires, self-assembled resonators, and active anisotropic matter. However, 
progress in this direction is hindered by a lack of control of the kinetics and microstructure due to 
inherent complexity arising from competing energy and topology. We have studied thermal and electrokinetic 
effects on disclinations in a three-dimensional nonabsorbing nematic material with a positive and negative 
sign of the dielectric anisotropy. The electric flux lines are highly non-uniform in uniaxial media after 
an electric field below the Fr\'{e}edericksz threshold is switched on, and the kinetics of the disclination 
lines is slowed down. In biaxial media, depending on the sign of the dielectric anisotropy, apart from 
the slowing down of the disclination kinetics, a non-uniform electric field filters out disclinations 
of different topology by inducing a kinetic asymmetry. These results enhance the current understanding 
of forced disclination networks and establish the presented method, which we call fluctuating 
electronematics, as a  potentially useful tool for designing materials with novel properties \it{in silico}.
\end{abstract}

\maketitle
\section{\sc \bf Introduction}
\label{sec:1}

Topological singularities such as points, lines and walls are ubiquitous in phases with broken symmetry. 
Canonical examples include dislocations in solids\cite{klelav}, vortex lines and rings\cite{rututu} in 
superfluid ${}^3He$ and ${}^4He$, Abrikosov vortex lines in superconductors\cite{bigahumur}, vortex 
lines in Bose-Einstein condensates\cite{weneschbrdaan}, umbilic lines\cite{macgar} and $\pi$ 
solitons\cite{veinblsmlavnob} (disclinations) in nematic liquid crystals (NLC) that provide a testing 
ground for theories of cosmology\cite{chdutuyur,rututu}, $\lambda$ lines in cholesteric fluids\cite{klefrie}, 
Bloch and N\'{e}el lines in ferromagnets\cite{kleman}, walls in lipid membranes\cite{drbrjoad} and 
string networks in ecology\cite{avbamenoli}. NLC phases display rich birefringence under a polarizing 
microscope during phase ordering from a disordered state after a rapid quench in pressure or temperature, 
resulting in the formation of disclinations with integer and fractional topological charge. These 
singularities proliferate after nucleation and form contractile loops after intercommutation\cite{mermin}. 
Unlike dislocations, disclinations possess intricate kinetics, microstructure, and equivalence with 
an electric charge. Strings are charge neutral with either topological charge $\pm1$ or $\pm1/2$ 
residing at the two segments or end points of the string to form topological dipoles. Higher multipoles 
and integer charged dipoles also nucleate within the charge neutral strings at the early stage of 
kinetics. Subsequently, these structures rupture into fractionally charged dipolar strings. Similar 
to electrodynamics, like topological charges repel and unlike charges attract and annihilate in pairs 
while monopoles are nonexistent to retain charge neutrality unless created by symmetry-breaking 
boundaries, an inclusion of impurity or external drive with a laser beam\cite{niskcorazumu}. 

\begin{figure*}
\centering
\includegraphics[width=0.75\textwidth, height=0.6\textwidth]{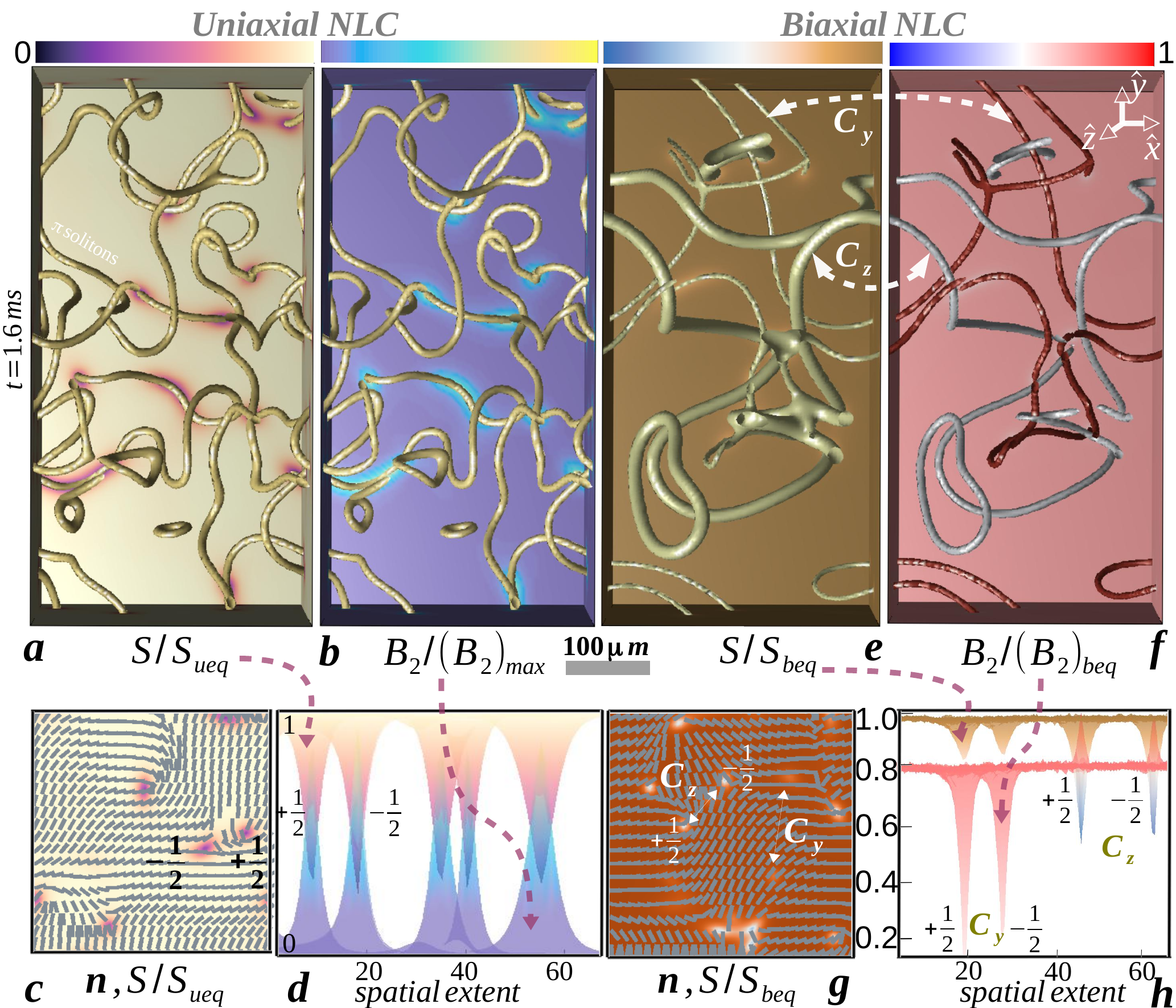}
\caption{{\bf a,} $\pi$ solitons in a thick uniaxial film of $5CB$ after a temperature quench from an isotropic 
state ($T=40^{\circ}$C) to a nematic state ($T=33.65^{\circ}$C). Supercooling and superheating temperatures 
are $T=\{34.2,34.47\}^{\circ}$C. Disclination isosurfaces correspond to scalar uniaxial order 
with isovalue $S_{ueq}/2$, where $S_{ueq}=0.086$. {\bf b,} Corresponding biaxial order with isovalue 
$(B_2)_{max}/2$ where $(B_2)_{max}=0.05$. {\bf c,} uniaxial order and director distribution on a portion 
of the ${\it xy}$-slice plane of ({\bf a}). {\bf d,} Corresponding spatial extension of scalar uniaxial 
and biaxial order, displaying core structure of two segments of charge neutral disclinations in-plane, 
which are $\pm1/2$ integer defects\cite{veinblsmlavnob,niskcorazumu}. {\bf e,} Charge neutral $\pi$ 
solitons of different homotopy class in thermotropic biaxial media (see Supplementary Movie S2 and 
Appendix (section \ref{app:3}) for defect characterization scheme). Disclination isosurfaces correspond to isovalue $6S_{beq}/7$ 
where $S_{beq}=0.96$. {\bf f,} Corresponding biaxial order with isovalues \{$0.42(B_2)_{beq}$,$0.83(B_2)_{beq}$\} 
for homotopy class $\{C_y,C_z\}$ where $(B_2)_{beq}=1.2$. {\bf g,} Uniaxial order and director distribution 
on a portion of the ${\it xy}$-slice plane of ({\bf e}). Note the similarity in the microstructure of 
$\pm1/2\;C_z$ defects with uniaxial defects in ({\bf c}) and no variation in {\bf n} for $\pm1/2\;C_y$ 
defects. {\bf h,} The corresponding variation of uniaxial and biaxial order displaying core structure 
of disclination segment of both class in-plane. Although both scalar orders decrease in the core of 
$C_y$ defect, biaxiality increases for a decreasing uniaxiality in $C_z$ defect. Parameters are defined 
in Appendix (section \ref{app:1}) and material (computation) parameters are tabulated in Table \ref{tbl:GLdGparam}.} 
\label{fig:1}
\end{figure*}

Existence, classification and recombination rules of disclinations in equilibrium, which play an 
important role in the material design, is governed by the energy landscape as well as the geometry 
(topology) of the order parameter space\cite{mermin}. Strings in uniaxial NLC displayed in figure
\ref{fig:1} (frames {\bf a-d}) are topologically defined by $\pi_1(\mathcal{RP}_2)=\mathbb{Z}_2$ 
with homotopy group $\pi_1$ in the projective plane $\mathcal{RP}_2$ resulting in the abelian group 
$\mathbb{Z}_2$ with topological charge $\pm1/2$\cite{mermin}. After a theoretical proposal\cite{frank}, 
$\pi$ solitons have been seen in fluorescence confocal polarized-light microscopy of pentylcyanobiphenyl 
(5CB) NLC\cite{chdutuyur,veinblsmlavnob}, molecular simulations\cite{bismallopel} and field theoretic 
computations\cite{bhatam,niskcorazumu}. Likewise, biaxial disclinations displayed in figure \ref{fig:1} 
(frames {\bf e-h}) are defined by $\pi_1(\mathcal{RP}_3)= \mathbb{Q}_8$ where $\mathcal{RP}_3$ is the 
projective plane and $\mathbb{Q}_8$ is the nonabelian group of quaternions generating three classes 
of half-integer topological charges denoted by $C_{x,y,z}$. In monolayered thin films, the simultaneous 
and pairwise coexistence of fractionally charged point dipoles of either class $C_{x,y},C_{y,z}$ or 
$C_{x,z}$ is predicted\cite{kobtho} and observed in field theoretic computations\cite{zapgolgol,bhatam}. 
Albeit topologically proscribed in three dimensions, strings of disparate topology do not entangle but 
annihilate pairwise within the respective class\cite{bhatam}.
Depending on the anisotropic elastic constants of the medium, soft disclinations are vulnerable to 
thermal fluctuations and external stimulus like an electromagnetic field. Regulated by the sign of 
the dielectric anisotropy constant of the material, an electric field at the Fr{\'e}edericksz threshold 
can orient the nematic director along or perpendicular to the field direction. It is particularly 
interesting to examine whether locally uniform and nonuniform electric field can lead to a time-dilated 
kinetics of the disclination network\cite{niskcorazumu}, and, how the anisotropy of the nematic 
orientation embedded in the dielectric tensor leads to nonuniformity in the local electric field.
For example, nematic regions at the top of a colloidal inclusion are generated due to the asymmetric 
distribution of the field intensity\cite{ucaronu}. Such control is hard to characterize in experiments, 
impossible in nanoscale molecular simulations and limited in field theoretic calculations due to 
numerical complexity, as Ref.\cite{ucaronu} mentions, ``molecular alignment in the inhomogeneous 
electric field has not yet been well studied as it is not easy to solve the Poisson equation with 
an inhomogeneous dielectric constant to calculate the local electric field''. Attributing to the 
scale invariant property of the Ginzburg-Landau-de Gennes (GLdG) field theory, relaxational kinetics 
of the orientation tensor has quantitatively reproduced experiments {\it in silico} from 
mesoscale\cite{bhatnucl,niskcorazumu} to nanoscale\cite{winalej}. {\it State of the art} grand 
challenge is attributed to the nonavailability of a robust numerical scheme\cite{abukhdeir} which 
is, in descending order of complexity, (a) free from numerical artifacts of the traditional 
methods\cite{todeyeo}, accounts for (b) local nonuniformity in electric field and (c) equilibrium 
thermal fluctuations by respecting physical laws, (d) guarantees zero-trace property of the 
orientation tensor and (e) incorporates anisotropic elasticity to probe beyond the single diffusion 
(one elastic constant) approximation\cite{todeyeo}. Recent advances in fluctuating hydrodynamics of 
isotropic suspensions\cite{donobhgabe} incorporating point (a) demand a natural, yet challenging, 
extension for anisotropic suspensions\cite{leonar} while on the other hand, numerical achievement 
of points (c-e) is fairly recent\cite{bhmeads,bhatnucl}.
\begin{table*}
\small
\begin{tabular*}{1.0\textwidth}{@{\extracolsep{\fill}}||c||c||c||c||c||c||c||c||c||c||c||c|}
\hline
Fig. & $\Gamma (P^{-1})$ & $A (Jcm^{-3})$ & $B(Jcm^{-3})$ & $C(Jcm^{-3})$ & $E^\prime(Jcm^{-3})$ 
&$L_1 (10^{-7}dyn)$ & $\kappa$ & $\Theta$ & $\zeta (\mu m)$ & $k_BT (J)$ \\
\hline
\hline
1{\bf a-d} & $5$ & $$ & $$ & $$ & $$ & $3.75\times10^{-3}$ & $18$ & $0.5$ & $$ & $0$\\
\cline{11-11} 
\ref{fig:3}      & $$  & $$ & $$ & $$ & $$ & $$             & $$   & $$    & $$ & $$\\
\cline{2-2} \cline{7-9}
\ref{fig:2}{\bf a-b},\;\ref{fig:4} & $1$ & $-8\times10^{-3}$ & $-0.5$ & $2.67$ & $0$  & $0.05$ & $0$ & $0$ & $3.55$ & $5\times10^{-6}$ \\
\cline{2-2} \cline{7-9}
$$     & $$    & $$       & $$     & $$     & $$   & $0.05$ & $0$  & $0$   & $$   & $$\\
\ref{fig:2}{\bf c}      & $5$    & $$ & $$ & $$ & $$  & $6.8\times10^{-3}$  & $9$  & $0.5$ & $$     & $$ \\
$$      & $$    & $$       & $$     & $$     & $$   & $3.75\times10^{-3}$ & $18$ & $0.5$ & $$ & $$\\
\hline
\hline
\ref{fig:1}{\bf e-h},\;\ref{fig:S1} & $0.02$ & $-4.5$ & $-0.5$ & $2.67$ & $3.56$ & $8.1$  & $0$  & $0$   & $3.55$ & $0$\\
{\ref{fig:5},\;\ref{fig:6}}            & $$   & $$   & $$   & $$   & $$   & $$ & $$ & $$  & $$   & $8\times10^{-3}$\\

\hline
\end{tabular*}
\caption{\label{tbl:GLdGparam} Parameters values excercised to mimic uniaxial (upper row) and biaxial 
(lower row) thermotropic NLC. A rectangular simulation box of size $80^2\times160{\mu}m^3$ with grid 
spacing $\Delta x = \Delta y= \Delta z= 1{\mu}m$ and time step $\Delta t=1{\mu}s$ is considered. We 
use material parameters for $5CB$ at $T=33.65^{\circ}$C, $\epsilon_0=1,\epsilon_a=\pm1, 
\epsilon_s=0.74\epsilon_a$\cite{colhird} and use earlier excercised material parameters for biaxial 
media\cite{gralondej,bhatam}. Using equation (\ref{eq:3}), we estimate $E_F=2\times10^{-3} V/{\mu}m$ 
for $5CB$ and with $|E|=E_F\times(10^{-2},10^{-1},1)$, nondimensional parameters are $\epsilon_1=
1.5\times(10^{-4},10^{-3},10^{-2})$, $\epsilon_2=1.8\times(10^{-3},10^{-2},10^{-1})$. In biaxial 
media, we use $|E|=1.5 V/{\mu}m<E_F, \epsilon_1=0.14, \epsilon_2=3.35$. $t_{on}=1ms$ for both uniaxial 
and biaxial problem, while $t_{off}=5ms$ for uniaxial and $t_{off}=\infty$ for biaxial problem. 
Parameters are defined in Appendix (section \ref{app:1}).}
\end{table*}

By investigating beyond the uniform field assumption\cite{olavmool}, in this article, we have developed 
a fluctuating electronematics method based on the thermal description of the GLdG theory, with the physical 
control over the role of each forcing to accurately describe thermal and electrokinetic phenomena in three 
dimensional NLC media. Using simple analytical argument, we provide an understanding 
of the underlying mechanism responsible for the outcome in both uniaxial and biaxial media in the free 
draining limit at moderate to small electric field intensity, where the advective flow of the anisotropic 
media can be neglected. The interdependency between the topology in the orientational order of the 
media and morphology of the external field is elucidated through the measurement of disclination kinetics 
and morphology. It turns out that small magnitude of nonuniform electric field can significantly dilate 
the coarsening kinetics of disclination network and can eradicate certain topological class of disclinations 
in biaxial media.
\begin{figure*}
\centering
\includegraphics[width=1.0\textwidth, height=0.4\textwidth]{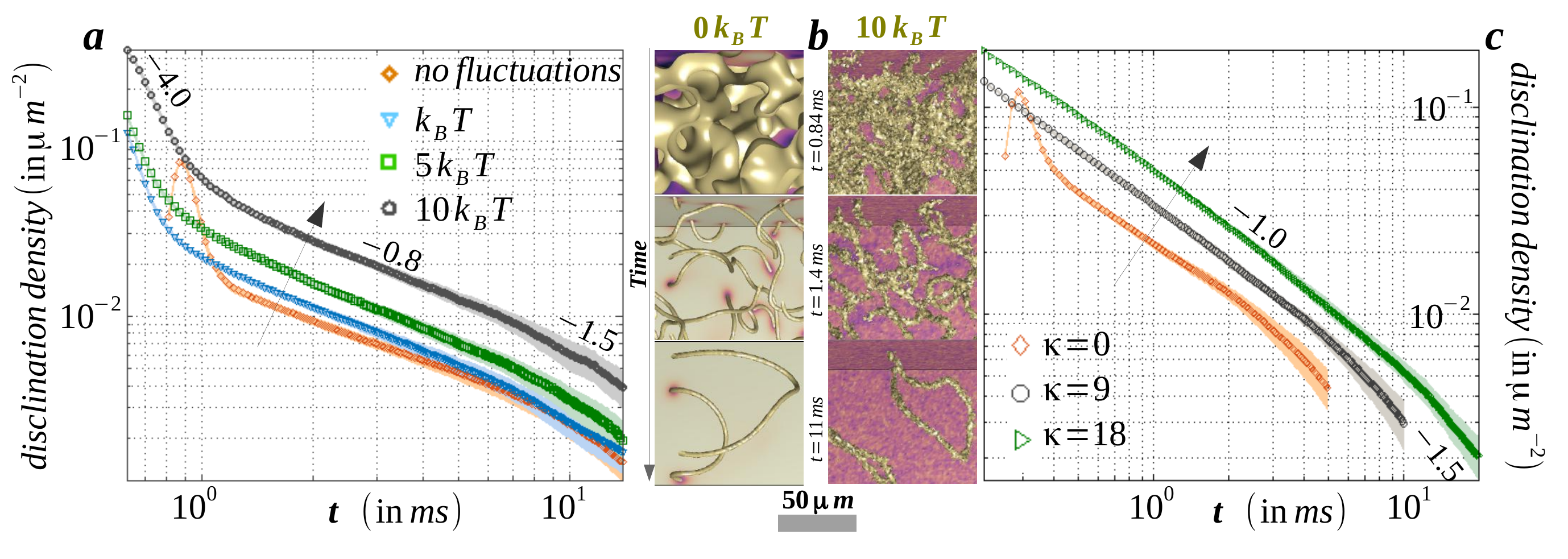}
\caption{\label{fig:2}{\bf a,} Increment in disclination density with fluctuation amplitude. Note the 
three different slopes corresponding to initial diffusive regime $\to$ Porod's law regime $\to$ late stage 
diffusive regime. {\bf b,} Comparison of disclination kinetics in a thermally fluctuating media with its 
athermal variant that corresponds to ({\bf a}). A portion of the three-dimensional volume is shown for 
clarity, that displays domain decomposition before nucleation of disclinations $\to$ intercommuting 
disclinations $\to$ contractile disclination loops. {\bf c,} Increased elastic anisotropy leads to 
increased disclination density with a prolonged Porod's law regime. Note that both thermal fluctuations 
and elastic anisotropy aids in early nucleation of the isotropic domain, but prolongs the disclination 
kinetics. The rendered colours in field values are indicated in figure \ref{fig:1} and the arrow denotes 
the increment direction. Material (computation) parameters are tabulated in Table \ref{tbl:GLdGparam}.} 
\end{figure*}

\begin{figure*}
\centering
\includegraphics[width=1.0\textwidth, height=0.75\textwidth]{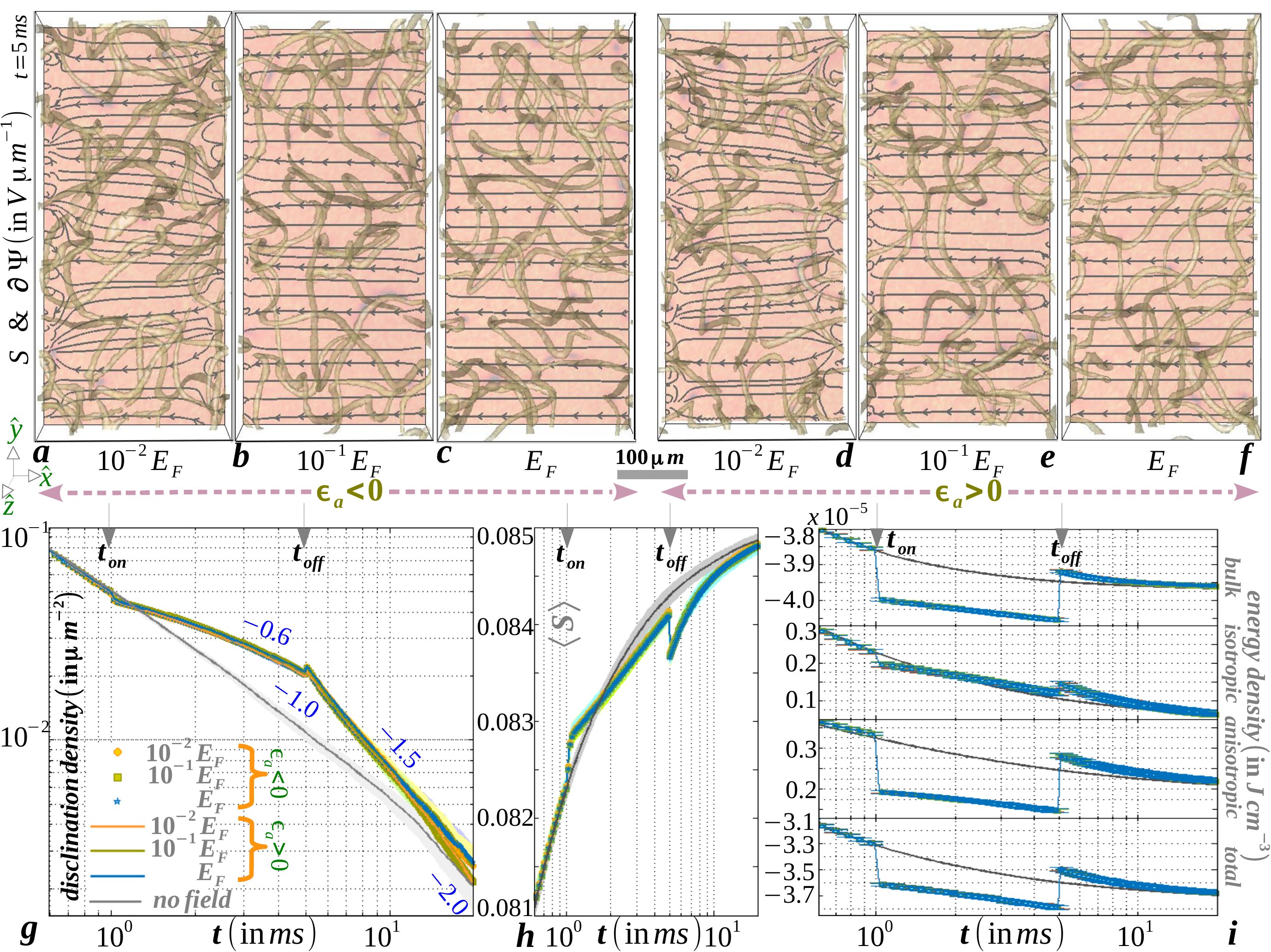}
\caption{\label{fig:3}{\bf a-c,} Nonuniformity of electric flux lines along with the uniaxial order 
is sketched on slice plane $L_z/8$ for negative dielectric anisotropy constant material (for example, 
butylmethoxybenzylidene ($MBBA$)) when the magnitude of the non-dimensionalized scalar potential is 
comparable to the orientation tensor (see Supplementary Movie S3). Disclinations in the bulk volume 
are also displayed to guide the eye. Uniformity in flux lines is gradually reached as the field is 
increased to the Fr\'{e}edericksz threshold $E_F$. {\bf d-f,} Comparatively similar electric response 
of a positive dielectric anisotropy constant material (for example, $5CB$). {\bf g,} Evolution of 
disclination density per unit area displaying an increased life-span of $\pi$ solitons with non-Markovian 
response during field cessation, {\bf h,} elastic response of the global uniaxial order and {\bf i,} 
(free) energy with varying field strength after onset and cessation of an unidirectional electric 
field. Equilibrium response is also sketched for comparison. Material (computation) parameters are 
tabulated in Table \ref{tbl:GLdGparam}.} 
\end{figure*}

\section{\sc \bf Results}
\label{sec:3}
\subsection{\sc \bf Thermoelectrokinetic effects of disclination network in a coarsening uniaxial NLC}
\label{subsec:1} 
Here we systematically probe on the role played by the elasticity of the medium and various external 
forcing ($k_BT, {\bf E}$) on the kinetics and microstructure of the string disclination assembly.
 
{\sc \bf Role of thermal fluctuations and anisotropic elasticity.}
The disclination kinetics is immensely influenced by external agents like thermal and electric forces, 
anisotropic elastic effects or shear. To illustrate the role of thermal fluctuations, we estimate the 
decay of disclination density per unit area in the degenerate elastic constant approximation without 
an electric field ($\kappa=\Theta=|E|=0$ in equation (\ref{eq:6})). Using surface triangulation 
method\cite{hjedae}, we calculate the total surface area of the disclinations in the media. Figure 
\ref{fig:2}: frame {\bf a} plots the same for three different values of $k_BT$ including the athermal 
scenario and in frame {\bf b}, we show a portion of the three-dimensional volume that distinguishes 
the disclination kinetics between the athermal and thermal scenario. In the athermal scenario, three 
different regime with marked exponents emerges out in the evolution process, that had been previously 
quantified as diffusive regime $\to$ Porod's law regime $\to$ diffusive regime\cite{bhatam}. The early 
diffusive regime corresponds to domain coarsening before nucleation of disclinations ($t=0.84ms$), while 
Porod's law scale designates the defect annihilation kinetics ($t=1.4ms$) and finally, the late stage 
diffusion is attributed to contraction of isolated loops ($t=11ms$). Clearly, thermal fluctuations tend 
to increase the disclination surface density without affecting the scaling laws, that is important for 
materials ({\it e.g.} PAA) having transition temperature way above the room temperature. Fluid viscosity 
$\eta$ can be obtained from the Stokes-Einstein relation $k_BT/K\eta$ = constant. However, frame {\bf c} 
shows that disclination density in the athermal media not only increases for an increase in $\kappa$, but 
also  stretches the Porod's law regime. The slope changes from $0.8$ to $1$ as elastic anisotropy is 
increased. While the slope of unity is also obtained in experiments with $5CB$\cite{chdutuyur,chtuyur}, 
we re-establish our earlier claim\cite{bhatnucl} about the crucial contribution of the anisotropic 
elasticity of the medium. Intuitively, anisotropic elastic constant results into asymmetric diffusion 
constants in Cartesian directions and thus brings asymmetry in the speed of $\pm1/2$ integer point 
defects\cite{todeyeo,niskmus}. The loss of area in forming a contractile loop is greatly reduced for 
higher anisotropy.

Before we embark on discussing coarsening in the presence of an electric field, we revisit the coarsening 
kinetics when no electric field is present. As seen in Supplementary Movie S1, disclinations with higher 
line tension and curvature are energetically disfavoured, culminating in stretched strings that form 
contractile loops after intercommuting with neighbouring strings. By taking into account the effect 
of viscous drag and length decrement of a string in forming a loop at time $t$, the disclination 
surface density $\rho$ is found\cite{chdutuyur} to scale as $\rho\propto t^{-1}$. An identical scaling 
law is obtained for a planar disclination when equating the rate of change of the line tension per unit 
volume with the energy density loss rate\cite{klelav}. Our method accurately reproduces the slope of 
$1\pm10^{-3}$ within the errorbar shown in figure \ref{fig:3} (`no field' curve in frame {\bf g}) 
with the material parameters of $5CB$. Thermal fluctuations and elastic anisotropy lead to an increment 
in the disclination surface density at a given time and the period of disclination annihilation kinetics 
is extended without affecting the physical laws. 

{\sc \bf Role of an electric field.}
Next, we elaborate on the thermal and electrokinetic effects on coarsening in uniaxial NLC after the onset 
and cessation of an unidirectional voltage pulse. The electric field is applied at an instant $t_{on}$ when 
only disclinations with dipolar charge $\pm1/2$ are present, and, the medium is free of $\pm1$ dipolar strings 
or point charges. The switch-off time of the field is set at an instant $t_{off}$ when the electric flux 
lines within the medium do not  change substantially. For a shallow quench below the supercooling line, for 
both signs of the dielectric anisotropy constant and below the Fr\'{e}edericksz threshold of the electric 
field, the orientational order is comparable in magnitude with the non-dimensionalized electric potential. 
Thus within the medium, the flux lines are very much distorted as seen in figure \ref{fig:3}. 
Instead of being aligned along the field direction, the isotropic cores of the strings displaying reduced 
uniaxial order are little deformed by the electric force\cite{veinblsmlavnob} and have a little 
contribution in distorting the flux lines. This is attributed to the weak coupling of scalar order to the 
electric field, unlike the director that strongly couples to the electric field. In no switch-off scenario 
($t_{off}\to\infty$), the nonuniformity of the flux lines is retained in the electrically forced nematic 
phase devoid of disclinations around $t\sim35ms$ (not shown). The flux lines in figure \ref{fig:3} (frames 
{\bf a-f}), however, tend to gain uniformity as the electric field is increased towards the Fr\'{e}edericksz
threshold for both signs of the dielectric anisotropy constant. As observed in frame {\bf g}, 
electrically forced strings are  little thinner due to a reduction of the surface density, but they are 
long lived due to a dilated kinetics. After the field is switched off, disclination 
kinetics and surface density do not immediately return to the zero-field behavior, but lag for an interval 
of $\sim 10ms$. Thus, the material retains a memory of the field onset in the process of exhibiting an 
elastic response. The electric field agitates the isotropic background towards an uniaxial medium, thus 
increasing the uniaxial order (frame {\bf h}) and non-monotonically decreasing the total free 
energy (frame {\bf i}). Though we obtain Brochard-Leg\'{e}r lines connecting $\pi$ solitons under 
an intense field above the Fr\'{e}edericksz threshold, the lack of backflow in our method cannot reproduce 
a ceasing motility of disclinations\cite{veinblsmlavnob}. Rather, they annihilate at a faster rate due to 
the uncompensated electric drag force. Electrokinetic effects under intense forcing, such as electroconvection 
or rheochaos, can be quantitatively captured only if backflow is systematically included. Unlike in colloidal 
suspensions\cite{donobhgabe}, the question of the existence of correlations between fluctuations in orientation 
and velocity has to be answered from experiments\cite{zarsen} before attempting a numerical study in which 
backflow effects are included.
\begin{figure}[t]
\centering\includegraphics[width=0.45\textwidth, height=0.35\textwidth]{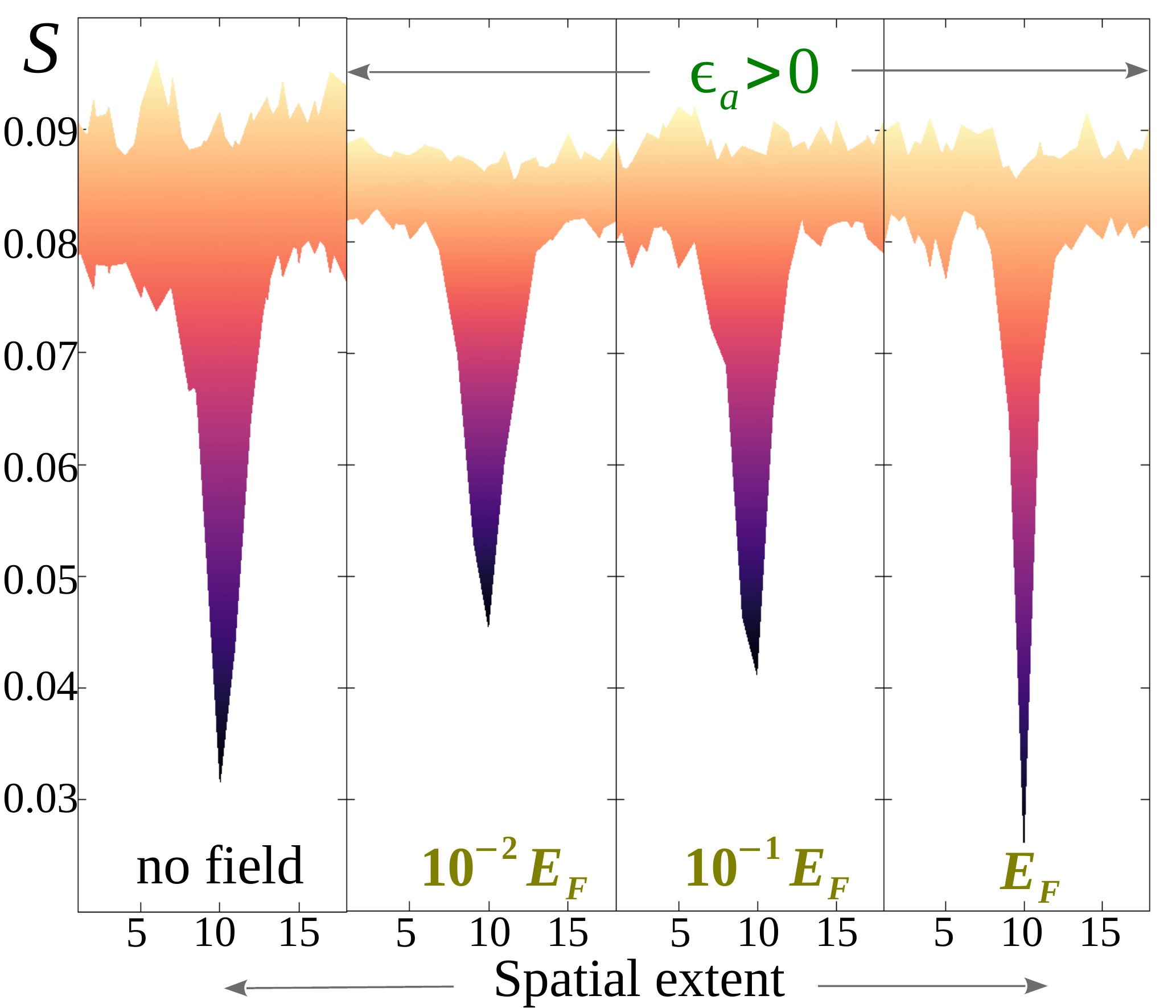}
\caption{\label{fig:4} Spatial extent of surfaces of isolated uniaxial planar disclinations obtained from 
the portion of the ${\it xy}$-slice plane of the three-dimensional volume in figure \ref{fig:2} (frame {\bf 
b}). Colourbars are indicated in figure \ref{fig:1}. Note the reduced fluctuation amplitude of $S$ for 
$|E|\neq0$ compared to no field defect core.}
\end{figure}

We qualitatively argue about the slowing down of disclination kinetics for a planar isolated disclination 
loop. An in-plane estimate is valid for $5CB$ while twist constant is much smaller than splay or bend 
constants\cite{bhatnucl}. Also deep within the uniaxial phase where the director is aligned uniformly, 
Frank constant $K$ can sufficiently define the medium elasticity\cite{bhmeads}. In such situations, planar 
disclination energy per unit length is $\mathcal{F}_{discl}=\int_{\mathbb{S}} d^2 x[K(\pmb\partial f)^2 - 
\epsilon_0\epsilon_a E^2 sin^2f]/2$ where $f=k\phi+c$, with $0\le\phi\le2\pi,\; k=\pm1/2$ being the 
topological charge, $\mathbb{S}$ the bounding plane and $c$ a constant, defines a planar defect 
configuration ${\bf n}=[cosf,sinf]$. Performing the surface integral, the reduced disclination energy 
per unit length is
\begin{equation}
\label{eq:7}
\mathcal{F}_{discl}=\pi Kk^2ln\Big({\xi \over \zeta}\Big) - {\pi\epsilon_0\epsilon_a \over 4k} 
E^2(\xi^2-\zeta^2) + \mathcal{F}_{discl}^{c}, 
\end{equation}
where $\mathcal{F}_{discl}^{c}$ is the disclination core energy. Equating the elastic energy 
$\mathcal{F}_{discl}/\xi$ per unit area with the drag force $-\eta\partial_t{\xi}$ per unit length 
yields contributions from (i) the equilibrium kinetics $\xi\sim t^{-1/2}$ and (ii) the electric field 
effect $\xi\sim e^{\nu t}$ with $\nu=\pi\epsilon_0\epsilon_aE^2/8k\eta$. While $\nu$ is independent 
of the sign of $k$ or $\epsilon_a$, $\nu>0$ implies of a temporal reduction of the loop extinction 
kinetics. Physically this can be interpreted as a reduction in speed of approach between $\pm1/2$-charged 
topological dipole within a charge-neutral loop due to the external forcing. As observed in figure 
\ref{fig:3} (frame {\bf g}) and in supplementary figure S1, nonuniform electric field substantially 
prolong the kinetics when compared to the uniform field scenario (see Supplementary Movie S4). 

To shed light on the effect of electric forces on the disclination core structure in thermal uniaxial 
media with $\epsilon_a>0$, in figure \ref{fig:4} we sketch the spatial variation of the surface of $S$ 
around a planar defect for different field intensity and compare with the equilibrium scenario. The 
fluctuation amplitude at $S_{ueq}$ is reduced due to the application of an electric field, resulting 
in a reduction of the disclination surface density (figure \ref{fig:3}, frame {\bf g}). However, we do 
not find any significant distortion of the core for an increasing field strength which indicates that 
the field, unlike the director orientation, cannot influence the sufficiently isotropic core other 
than a complete melting of the disclination at $E\gg E_F$.

\subsection{\sc \bf Thermoelectrokinetic effects in coarsening biaxial NLC.}
Next, we examine the role played by the isotropic elasticity of the medium and various external forcing 
($k_BT,{\bf E}$) on the kinetics and microstructure of the string disclination assembly of different 
homotopy class. We do not find excitingly different outcome when investigating the role of anisotropic 
elasticity and, thus, here we restrict ourselves in reporting results in $\kappa=0$ limit, in par with 
other investigations\cite{gralondej,bhatam}. 
\begin{figure}[t]
\centering
\includegraphics[width=0.5\textwidth, height=0.4\textwidth]{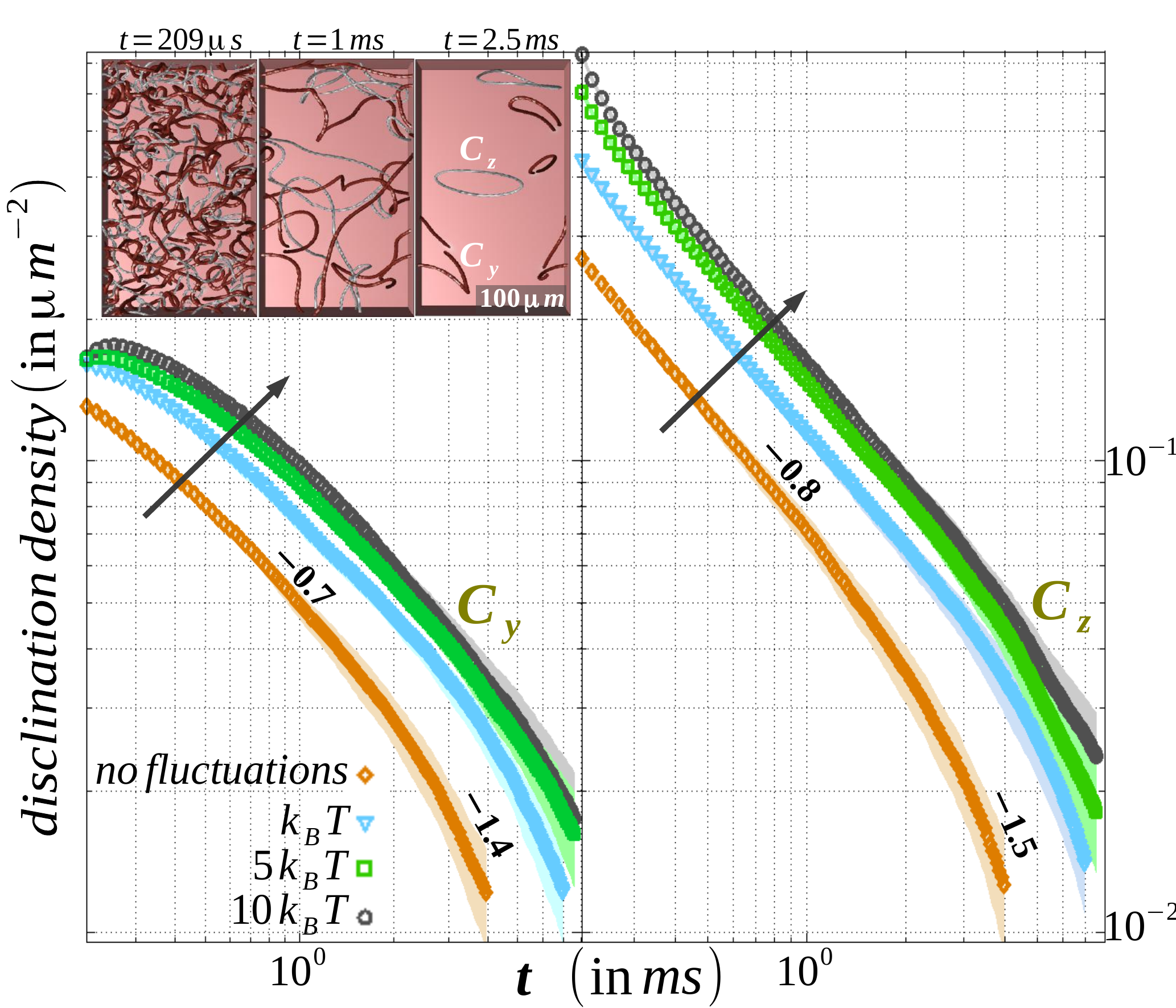}
\caption{\label{fig:5} Evolution 
of disclination density of $C_y$ class (left panel) and $C_z$ class (right panel) with fluctuation amplitude 
for biaxial nematic media is displayed. The direction of the arrow shows the increment of surface density with 
fluctuation amplitude. In the inset to the left panel, biaxial order $B_2$ is sketched in which nucleation, 
intercommutation, and extinction by ring formation of disclinations is portrayed for clarity. Slope and 
errorbar are indicated within the graphics. Material (computation) parameters are tabulated in Table \ref{tbl:GLdGparam}.} 
\end{figure}

{\sc \bf Role of thermal fluctuations.}
We estimate the consequence of thermal fluctuations on the biaxial disclinations of class \{$C_y,C_z$\}. 
In figure \ref{fig:5}, we plot the evolution of disclination density for different values of $k_BT$ and 
compare with the athermal scenario. To remind, in the inset to the left panel, we portray the early, 
intermediate and late stage of the disclination kinetics. Similar to the uniaxial disclinations, here 
we also find that thermal fluctuation increases the disclination density per unit area with a comparable 
slope. The increase in slope for an increase in $k_BT$ during early stage of the kinetics for $C_y$ class 
hints for a delayed emergence of Porod's regime, that is absent in $C_z$ class. More interestingly, we 
observe an equivalence of the $C_z$ class of biaxial disclinations with that of the $\pm1/2$-integer 
disclinations in uniaxial nematics, both (i) in morphology (figure \ref{fig:1}: frame {\bf c} and frame 
{\bf g}) and (ii) kinetics, as observed in the identical slope of $0.8$ in the Porod's law scaling regime 
(figure \ref{fig:5}: right panel and figure \ref{fig:2}: frame {\bf a}). Also, the mismatch of the slope 
between the $C_y$ and $C_z$ class of disclinations (figure \ref{fig:5}) suggests of minor influence 
within each other in the course of annihilation.
\begin{figure*}
\centering
\includegraphics[width=1.0\textwidth, height=0.75\textwidth]{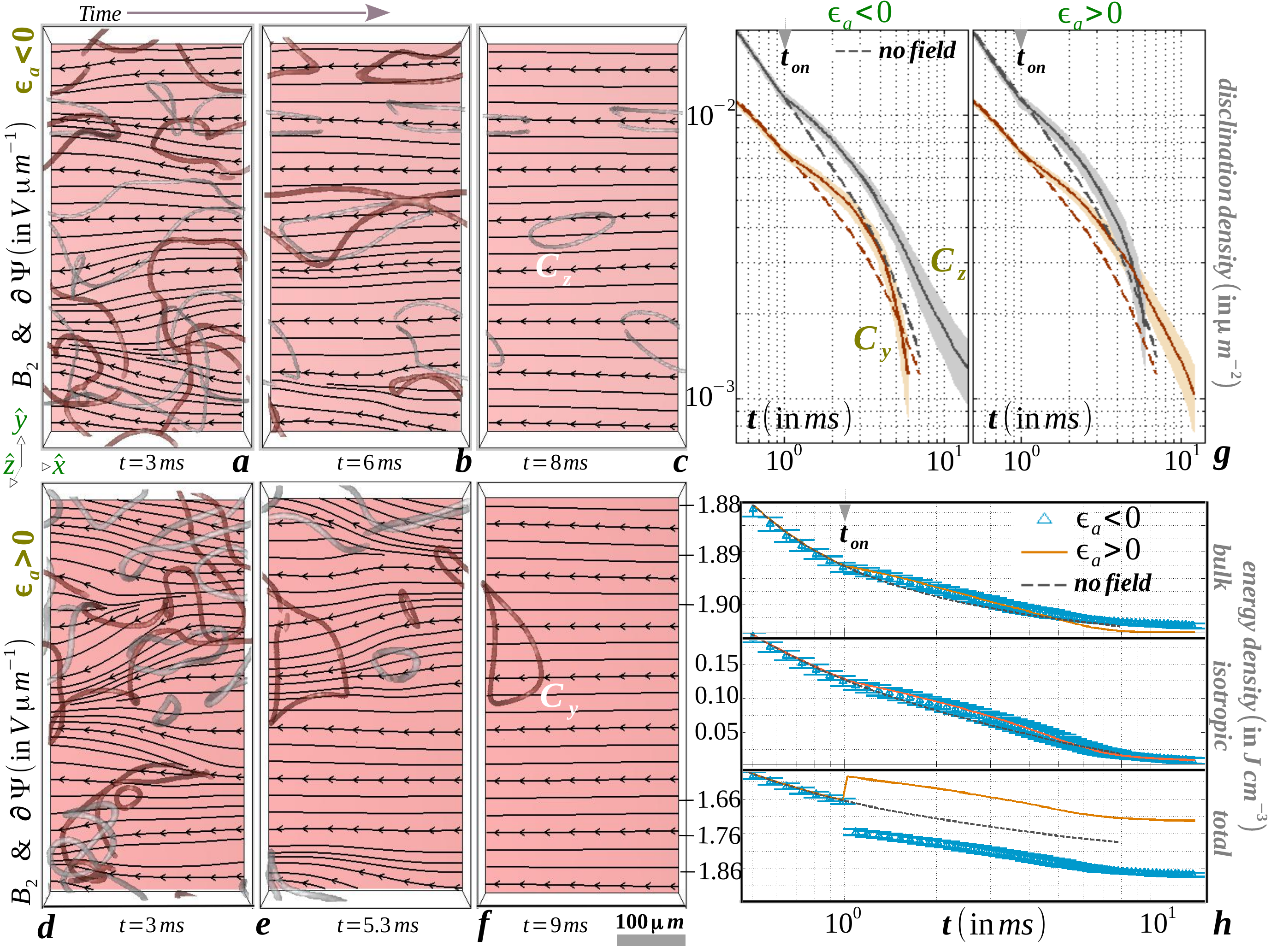}
\caption{\label{fig:6}{\bf a-c,} Evolution and selection of disclinations of homotopy class $C_z$ for a negative dielectric 
anisotropy constant material under the onset of an applied field $E=1.5V/{\mu}m$. The presence of $C_y$ 
disclinations at the early stage severely distorts the electric flux lines within the mesogen which regain 
uniformity as these disclinations are expelled from the medium (see Supplementary Movie S5-S6). {\bf d-f,} 
Similar response as previous for a positive dielectric anisotropy constant material, except that $C_y$ 
disclinations are selected and at an early stage, $C_z$ disclinations widely distort the electric flux 
lines which gain uniformity after their expulsion. {\bf g,} Decay of disclination surface density to 
portray quantitatively the selection of disclination class and {\bf h,} contributions from volume and 
surface energy to the total (free) energy of the medium. Equilibrium response is also sketched for 
comparison. Material (computation) parameters are tabulated in Table \ref{tbl:GLdGparam}.}
\end{figure*}

{\sc \bf Role of an electric field.}
To conclude, we examine the thermal and electrokinetic effects on the coarsening of biaxial NLC when the 
sample is rapidly cooled from a disordered phase in the presence of a steady voltage pulse. Figure 
\ref{fig:6} plots the instantaneous snapshots as well as the kinetic evolution of the medium. Although 
the topological structure and kinetic pathway of disclinations in biaxial NLC have been predicted 
for long\cite{zapgolgol,klelav,lucsluc}, experimental advance to stabilize disclinations by avoiding 
crystallization continues to be the {\it holy grail} of research on thermotropic biaxial 
mesogens\cite{madinasam}. Similar to the dilated kinetics of electrically forced uniaxial disclinations 
(see figure \ref{fig:3}), we find in figure \ref{fig:6} (frame {\bf g}) that the onset of an electric 
field increases the lifetime of biaxial disclinations of homotopy class \{${C_y,C_z}$\} at the initial 
stage for both sign of dielectric anisotropy. For either class of disclinations, 
$\nu(=\pi\epsilon_0\epsilon_aE^2/8k\eta)>0$ qualitatively explains the slowing down. For both signs of 
the dielectric anisotropy constant of the material, flux lines are massively distorted in the presence 
of disclinations (frames {\bf a,d}). After an interval of $\sim5ms$, a clear asymmetry between the 
disclination kinetics of different topology becomes evident. As shown in frames {\bf b-c,e-f,g}, this 
results in long-lived disclinations of either class with uniform electric field lines. This is attributed 
to an increment (decrement) of the total free energy with a positive (negative) value of the dielectric 
anisotropy constant (frame {\bf h}). Thus, the dielectric energy has a strong influence in selecting 
disclinations of the desired class as the bulk and elastic energies increase negligibly from the no 
field scenario. Physically, the acceleration (or retardation) in the loop extinction kinetics at the 
late stage can be interpreted as effective acceleration (or retardation) in speed of approach between 
$\pm1/2$-charged topological dipole within a charge-neutral loop of different class due to the electric 
force. Consistent evidence of class selection is also obtained, however on a much longer timescale, 
for values of the electric field magnitude much smaller than the Fr\'{e}edericksz threshold value. 
However, the decay kinetics of disclinations of a particular homotopy class is accelerated in the 
presence of an intense electric field due to the absence of backflow in our model to compensate the 
electric drag. We expect new phenomena in experiments on thermotropic biaxial media under an intense 
electric field, perhaps similar to the behavior of uniaxial disclinations under an intense electric 
field\cite{veinblsmlavnob}.

\section{\sc \bf Discussions}
\label{sec:4}
Electrorheology of line defects, with\cite{aratan,ravzum,limastca} or without\cite{todeyeo} 
particulate inclusion, under an intense electric field have established the coupling of orientation 
tensor with hydrodynamics and uniformity in the electric field, though the effect of nonuniformity in 
the electric field\cite{tofuaronu,foronu,ucaronu,cumecakon} and the effect of thermal fluctuations are 
less explored. The effect of hydrodynamics is assumed to be negligible under moderate to weak electric 
field intensity\cite{beredw}. We have examined the role of thermal fluctuations and nonuniformity 
of electric field in this limit and have shown that fluctuating electronematics is a robust tool 
to mimic laboratory experiments\cite{chdutuyur,veinblsmlavnob,niskcorazumu} {\it in silico} for 
anisotropic NLC in three dimensions. From the structure of the orientation tensor, we present a 
simple way to identify and classify the line defects and to compute physical quantities from the 
geometry of disclinations.

We have shown how the spatial uniformity in electric field is gained in approaching the Fr\'{e}edericksz 
limit. Apart from modifying the kinetic pathway of the coarsening of athermal disclination network, 
the external stimuli in terms of temperature fluctuations and local electric field essentially probe 
two emergent length scales: (i) interfacial correlation length between isotropic and nematic phase 
and (ii) radius of curvature of disclination loop. Neglecting any local heating effects due to the 
variation of temperature, any change in the correlation length is attributed to the disclination 
core size as well its geometric position within the three-dimensional volume. Although the evolution 
is temporally dilated compared to the no-field scenario, the external field cannot sufficiently modify 
the radius of curvature of isotropic disclinations - thus the lines are not stretched along the direction 
of the electric field, rather they retain their shape even when the director gets aligned along or 
perpendicular to the field direction depending on the sign of material's dielectric anisotropy. 
The inhomogeneity of the nematic orientation is manifest in the inherent nonuniformity of the local 
electric field - resulting in the highly nonuniform electric flux lines within the sample. The 
electric field induces a memory to the material that exhibit an elastic response and also induces 
a kinetic asymmetry within disclinations of the different class. On the other hand, increase in thermal 
fluctuations tends to increase the disclination surface density. 

This complex interaction can be intelligently engineered to yield a fascinating outcome in a more 
complex scenario, for example, fractal nematic colloids\cite{hajamoejmura}, metadevices\cite{zhekiv} 
and photonic applications\cite{obayya}. The electric field induced kinetic asymmetry leading to the 
class selection of biaxial disclinations can develop into novel materials in topologically similar 
systems. Other than NLC, the presented work has resemblance with line defects in passive\cite{rututu,bigahumur,kleman} 
and active\cite{bartolo,drbrjoad} soft matter including conducting microwires\cite{ralstan} and 
self-assembled resonators\cite{frashu}, and thus has the potential to bring exciting applications 
in diverse systems.

\section{\sc \bf Acknowledgements}
We thank S.Ramaswamy, and C.Dasgupta for a constructive criticism and careful reading a version of 
the manuscript. Including them, fruitful discussions with P.B.S.Kumar, R.Adhikari, and N.V.Madhusudana is 
gratefully acknowledged. We are thankful for a partial supercomputing support from {\it Thematic Unit of 
Excellence on Computational Materials Science} at Indian Institute of Science. This work is supported by 
the DST-INSPIRE grant number DST/04/2014/002537 of Govt. of India.

\appendix
\section{\sc \bf Fluctuating electronematics: model energy and thermal kinetics.}
\label{app:1}
Instantaneous orientational order, that distinguishes between the disordered liquid state and partially 
ordered nematic state, is characterized by a symmetric traceless second rank tensor\cite{klelav}
\begin{equation}
\label{eq:1} 
{\bf Q} = {1\over2}\big[3S\stl{\bf n\bf n} + B_2({\bf l\bf l} - {\bf m\bf m})\big]; \; 
-\frac{1}{3} \leq S \leq \frac{2}{3}, \; B_2 < 3S,
\end{equation} 
where \{{\bf n,l,m}\} denote the alignment direction of the long (director), intermediate (codirector) 
and short axis (secondary director) with the degree of uniaxiality $S \propto\langle Y_2^0\rangle_t$ 
and biaxiality $B_2\propto\langle Re[Y_2^{\pm2}]\rangle_t$. $Y_l^m$ are the spherical harmonics of 
degree $l$ with order $m$, $\langle\cdot\rangle_t$ denotes the ensemble average at that instant $t$ 
and symmetric traceless is symbolized with $\stl{\cdot}$. While $B_2$ cease to null as $S$ attains 
its maximum, other moments $R_2$ generated from the planar projection of \{{\bf n},{\bf l}\} is also 
exercised to define biaxiality\cite{lucsluc}. $B_2$ is numerically less expensive in considering one 
less degree of freedom, nevertheless, $R_2$ plays an important role in understanding the field-induced 
switching kinetics\cite{ribezan}. 

The ground state free energy including the excitations due to the spatial distortions and dielectric 
coupling are represented in the phenomenological Ginzburg-Landau-de Gennes (GLdG) free energy functional
\begin{equation}
\label{eq:2} 
\mathcal{F}_{total} = \int_\mathbb{V}d^3x (\mathcal{F}_{bulk}[{\bf Q}] + 
\mathcal{F}_{elastic}[{\pmb\partial \bf Q}] + \mathcal{F}_{dielec}[{\bf Q}]), 
\end{equation}
where $\mathbb{V}$ is the material volume. Bulk energy is superposition of absolute rotationally invariant 
functions of tensorial order $\mathcal{F}_{bulk}[{\bf Q}] = A\mathrm{Tr}{\bf Q}^{2}/2 + B\mathrm{Tr}{\bf Q}^{3}/3 + 
C(\mathrm{Tr}{\bf Q}^{2})^{2}/4 + E^\prime(\mathrm{Tr}{\bf Q}^{3})^{2}$, where parameters $\{A,B\}$ control 
the system temperature and size disparity, $C>0$ preserves the boundedness and $E^\prime\neq0$ brings the 
notion of biaxiality\cite{klelav}. Higher order expansions are not required while\cite{gralondej} 
$(\mathrm{Tr}{\bf Q}^{2})^{3}/6\geq(\mathrm{Tr}{\bf Q}^{3})^{2}$. Order in equilibrium is obtained from 
$\partial_S\mathcal{F}_{bulk}(S)=\partial_{B_2}\mathcal{F}_{bulk}(B_2)=0$, which for a uniaxial media is 
$S_{ueq} = -B/6C + ({B^2/36C^2}- {2A/3C})^{1/2};(B_2)_{ueq}=0$, while for a biaxial media, $(B_2)_{beq}= 
[-2A + S_{beq}(2B-3CS_{beq}+9E^\prime S_{beq}^3)]^{1/2}/[C+9E^\prime S_{beq}^2]^{1/2}$ and clumsy algebraic 
expression for $S_{beq}$ is omitted for brevity. 

Inhomogeneities due to the excitations above the ground state is concealed within the symmetry allowed 
lowest order terms in $\mathcal{F}_{elastic}[\pmb\partial{\bf Q}] = [L_{1} (\pmb\partial{\bf Q})^2 + 
L_{2}(\pmb\partial\cdot{\bf Q})^2 + L_{3}{\bf Q}\cdot({\pmb\partial{\bf Q}})^2]/2$, where ratio $\kappa=L_2/L_1$ 
and $\Theta=L_3/L_1$ of the elastic constants can be mapped to the Frank constants splay, bend and 
twist\cite{schtrim}. For $MBBA$, $(\kappa,\Theta)\approx1$ while for $5CB$, $\kappa\approx40,\Theta\approx1$ 
which designate nearly equal splay and bend constants for both materials ($\Theta\neq0$) but the twist 
constant is an order smaller than splay or bend. This results into nucleation of integer topological 
charged nematic droplets in the metastable isotropic medium of 5CB, while topologically uncharged 
nematic droplets nucleate in the metastable medium of MBBA\cite{chehamshe,bhatnucl}. 

The optical dielectric permittivity tensor is related to the orientation tensor when separated into 
symmetric and antisymmetric part $\pmb\epsilon = \epsilon_s\pmb\delta + \epsilon_a{\bf Q}$, where 
$\pmb\delta$ is the Kronecker delta, $\epsilon_s=\mathrm{Tr}{\pmb\epsilon}/3$ and 
$\epsilon_a=2(\epsilon_\parallel-\epsilon_\perp)/3$ with $\epsilon_\parallel(\epsilon_\perp)$ being 
the permittivity along (orthogonal to) the director. The material 
parameters are $\epsilon_s=0.74\epsilon_a$, where for $5CB$ $\epsilon_a=5.8$ at $T=33.65^\circ$C and 
$\epsilon_a=-0.7$ for $MBBA$ at $T=25^\circ$C\cite{colhird,blchig}. Application of a spatially varying 
electric field ${\bf E} = -\pmb\partial\Psi$ leads to an electric displacement ${\bf D} = \pmb\epsilon\cdot{\bf E}$ 
and therefore to a dielectric (free) energy term $\mathcal{F}_{dielec}[{\bf Q}] = - \epsilon_0{\bf D}\cdot\pmb\partial\Psi/8\pi$ 
with $\epsilon_0$ being the vacuum permittivity and $\Psi$ the electric potential. We characterize the 
intensity of the electric field with respect to the order\cite{ucaronu,admamiorl} and thermal energy by 
the nondimensional ratios $\varepsilon_1 = (\epsilon_0\epsilon_a E^2/8\pi AS_{ueq})^{1/2}$ and 
$\varepsilon_2 = (\epsilon_0\epsilon_a E^2/8\pi k_BT)^{1/2}$. We estimate the Fr\'{e}edericksz threshold 
to orient a director in a uniformly oriented nematic state along (orthogonal to) an electric field in a 
twist geometry, calculated by minimizing the free energy for director distortion and field coupling, to 
be\cite{freedric}
\begin{equation} 
\label{eq:3} 
E_F = \frac{\pi}{L_x}\Big[\frac{9S^2L_1\{1+\frac{2}{3}(\kappa+\Theta)\}}{2\epsilon_0\epsilon_a}\Big]^{1/2};\; 
\Psi_F = L_x E_F, 
\end{equation}
where $L_x$ is the spacing between the electrodes.

In the present study, confinement effects {\it e.g.} centrosymmetry breaking geometric restriction at the 
boundary by coverslips, patterned or chemically active walls, curvature induced polarization and the presence 
of free ions are not considered. For bent-core molecules, curvature induced polarization can be incorporated 
by adding $\pmb P_{f} = c_1 \pmb\partial\cdot{\bf Q} + c_2 \pmb\partial\cdot({\bf Q\cdot \bf Q}) + 
c_3 \pmb\partial(\text{Tr}{\bf Q}^2) + c_4[{\bf Q}\cdot (\pmb\partial\cdot{\bf Q})-({\bf Q}\cdot\pmb\partial)\cdot{\bf Q}]$ 
to the electric displacement {\bf D}, where $c_{1,\ldots,4}$ are coefficients\cite{blogama}. Free ions can 
also be neglected by retaining $n\ll \epsilon_0 \Psi_F/{eL_x^2}$, where $n$ is the free ion density and $e$ 
the electric charge\cite{ucaronu}.

When the medium is sufficiently {\it dry} so that the long ranged hydrodynamic interaction produced by 
the motile disclinations do not interfere the kinetics and fluid inertia plays no role - thus restricting 
to an overdamped relaxational kinetics without convection of momentum, the Langevin equation displaying the 
time evolution of the electric potential together with the orientation tensor can be written as
\begin{align}
\label{eq:4} 
\partial_t \Psi = {\pmb\partial}\cdot{\bf D};\; 
\partial_t {\bf Q} = - {{\pmb\Gamma}\cddot{\frac{\delta\mathcal{F}_{total}}{\delta {\bf Q}}}} + {\pmb\xi},
\end{align}
where ${\pmb\Gamma} = \Gamma\big[\delta_{ik}\delta_{jl}+\delta_{il}\delta_{jk}-2\delta_{ij}\delta_{kl}/3\big]$ 
is a $4^{th}$ rank tensor that maintains the symmetric-traceless property on the right-hand side of second 
equation (\ref{eq:4}). The rotational diffusion constant $\Gamma$ is approximated to be independent of {\bf Q} 
and the stochastic term ${\pmb \xi}$ satisfies the property of the orientation tensor $\langle\pmb\xi({\bf x}, 
t)\rangle = 0, \langle\pmb\xi({\bf x}, t)\pmb\xi({\bf x^{\prime}},t^{\prime})\rangle = 2k_BT 
{\pmb\Gamma}\delta({\bf x-x^{\prime}})\delta(t-t^{\prime})$ and is constructed as a summation of 
Wiener process to keep discrete fluctuation dissipation (FDT) spectrum intact over all Fourier modes 
and thus to sample Gibbs distribution in thermal equilibrium\cite{bhmeads}. We stress at this point that 
the functional derivative of $\mathcal{F}_{dielec}[{\bf Q}]$ with respect to $\Psi$ is the divergence
of the electric displacement {\bf D}\cite{tofuaronu}. This legitimate the imposition of Maxwell's equation 
along with the {\bf Q}-tensor equation. Unlike Ref.\cite{cumecakon}, we neglect cross-coupling between terms proportional to 
$\delta \mathcal{F}_{total}/\delta \psi$ to the right-hand side of ${\partial_t \bold Q}$ equation and 
$\delta \mathcal{F}_{total}/\delta {\bold Q}$ to the right-hand side of ${\partial_t \psi}$ equation for 
simplicity.

By substituting equation (\ref{eq:2}) in equation ({\ref{eq:4}}), the coupled Maxwell-GLdG equation 
in expanded form reads
\begin{align}
\label{eq:5}
\partial_t \Psi =& \epsilon_0(\epsilon_s\;\partial^2\Psi + \epsilon_a\partial_i Q_{ij}\partial_j\Psi), \nonumber \\
\partial_t Q_{ij} =& -\Gamma\big[(A + C\mathrm{Tr}{\bf Q}^2)Q_{ij} + (B + 6E^{\prime}\mathrm{Tr}{\bf Q}^3)
\stl{Q_{ij}^2} - \nonumber \\
& L_1\{\partial^2 Q_{ij} + \Theta\stl{(Q_{mn}\partial_m\partial_nQ_{ij}-{1\over2}\partial_iQ_{qr}\partial_jQ_{qr})} 
 + \nonumber \\
& \kappa\stl{\partial_i \partial_l Q_{jl}}\} - {1 \over {8\pi}}\epsilon_0\epsilon_a\stl{\partial_i\Psi\partial_j\Psi}\big] 
+ \xi_{ij}.
\end{align}
   
A significant departure from uniformity in electric flux lines is expected when $\mathcal{O}(\Psi/\Psi_F)
\sim\mathcal{O}({\bf Q})$. Thus, near and below the supercooling temperature $T<T^*$, the orientational order 
is small and is affected by a moderate to small magnitude of the electric force. Conversely, under an intense 
electric field $E\gg E_F$, $\mathcal{O}(\Psi/\Psi_F)\gg\mathcal{O}({\bf Q})$ and the electric field can 
remain spatially uniform by decoupling from the ${\bf Q}$-tensor equation (\ref{eq:5}) to yield, 
\begin{align}
\label{eq:6}
\partial_t Q_{ij} =& -\Gamma\big[(A + C\mathrm{Tr}{\bf Q}^2)Q_{ij} + (B + 6E^{\prime}\mathrm{Tr}{\bf Q}^3)
\stl{Q_{ij}^2} - \nonumber \\ 
& L_1\{\partial^2 Q_{ij} + \Theta\stl{(Q_{mn}\partial_m\partial_nQ_{ij}-{1\over2}\partial_iQ_{qr}\partial_jQ_{qr})}  
 + \nonumber \\
& \kappa\stl{\partial_i \partial_l Q_{jl}}\} - {1 \over {8\pi}}\epsilon_0\epsilon_a\stl{E_iE_j}\big] + \xi_{ij}.
\end{align}
An extensive numerical route of investigation is presented next. 

The deterministic part 
of the equation (\ref{eq:6}) has been widely exercised in two-dimensional monolayered thin films in one 
elastic constant approximation ($\kappa=\Theta=0$) to examine electrokinetic effects of point defects in 
switching experiments\cite{olavmool,avmoololi}. However, an impeccable role of backflow and elastic anisotropy 
is found to decipher asymmetric speed of $\pm1/2$ integer defects under intense electric field\cite{todeyeo,niskmus}, 
that also leads to ceasing motility of disclinations due to the backflow\cite{veinblsmlavnob}. A quantitative 
measure in three-dimensional media under intense electric field, albeit experimentally posed for single 
disclination in deep uniaxial state\cite{veinblsmlavnob}, is yet to be sought by including Beris-Edwards 
model\cite{beredw} for fluid flow to the presented fluctuating electronematics model. Here instead, we 
focus on moderate to the small magnitude of electric forces including thermal fluctuations at temperature 
close and below $T^*$ where advective effects can be neglected for simplicity. \\

\section{\sc \bf Stochastic method of lines for fluctuating electronematics.}
\label{app:2}
We consider a thick rectangular slab of insulating thermotropic NLC material in thermal equilibrium, with 
an $80{\mu}m\times160{\mu}m$ base and $80{\mu}m$ height. At equilibrium, periodic boundaries in three 
Cartesian directions are retained, that can be realized as a free standing anisotropic thick film from 
a groove. The electric potential at $x=0$ is fixed at zero and at $x=L_x$ is held according to the desired 
field magnitude with Dirichlet boundary condition and periodicity is maintained in the {\it yz}-directions. 
This is mimic by suspending the thick slab within two planar laser beams kept at a different potential or 
placing within an electrode without the notion of an {\it easy axis}.

An elegant way to numerically integrate equations (\ref{eq:5}-\ref{eq:6}) while retaining the 
symmetric-traceless property of \{${\bf Q,\pmb\xi}$\} tensor is by projection on a basis of five $3\times3$ 
matrices as\cite{kawehe,bhmeads} ${\bf Q} = \sum_l a_l {\bf T}_l; {\pmb \xi} = \sum_l \zeta_l {\bf T}_l$ 
$(l=1,\ldots,5)$, so that the kinetics is projected into the basis coefficients $a_l$. The fluctuating force 
in the projected basis has the property, $\langle \zeta_l({\bf x}, t) \rangle = 0, \langle \zeta_l({\bf x}, 
t)\zeta_m({\bf x}^{\prime}, t^{\prime}) \rangle = 2k_BT\Gamma \delta_{lm}\delta({\bf x - x^{\prime}})\delta(t 
- t^{\prime})$. After projection, equation (6) translates into
\begin{align}
\label{eq:S1}
\partial_t \Psi =& \epsilon_0(\epsilon_s\;\partial^2\Psi + \epsilon_aT_{ij}^l\partial_i a_l\partial_j\Psi), \nonumber\\
\partial_{t}a_l =& -\Gamma \Big[(A + C\mathrm{Tr}{\bf Q}^{2})a_l + (B + 6E^{\prime}\mathrm{Tr}{\bf Q}^3)
T_{ij}^l\stl{{Q_{ij}^{2}}} - \nonumber \\ 
& L_{1}\big\{\partial^{2}a_l + \Theta \big(Q_{mn}\partial_m\partial_na_l - T^l_{ij}\partial_ia_{p}\partial_ja_{p} \big) + \nonumber\\ 
& \kappa \stl{T^{l}_{ij}T^{p}_{jk}\partial_{i}\partial_{k}a_{p}}\big\} 
 - {1 \over {8\pi}}\epsilon_0\epsilon_a T_{ij}^l\stl{\partial_i\Psi\partial_j\Psi}\Big] + \zeta_l,
\end{align}
while equation (7) takes the form,
\begin{align}
\label{eq:S2}
\partial_{t}a_l =& - \Gamma \Big[(A + C\mathrm{Tr}{\bf Q}^{2})a_l + (B + 
6E^{\prime}\mathrm{Tr}{\bf Q}^3)T_{ij}^l\stl{{Q_{ij}^{2}}} - \nonumber \\
& L_{1}\big\{\partial^{2}a_l + \Theta \big(Q_{mn}\partial_m\partial_na_l - 
T^l_{ij}\partial_ia_{p}\partial_ja_{p} \big) \nonumber \\
& + \kappa \stl{T^{l}_{ij}T^{p}_{jk}\partial_{i} \partial_{k}a_{p}}\big\} 
 - {1 \over {8\pi}}\epsilon_0\epsilon_a T_{ij}^l\stl{E_iE_j}\Big] + \zeta_l. 
\end{align}

To numerically integrate the above equations (\ref{eq:S1} or \ref{eq:S2}), we adopt a central finite 
differencing to spatially discretize the Laplacian and mixed derivates. An explicit stochastic method 
of lines (SMOL) integrator is exercised for seamless temporal integration\cite{wilkie}.  SMOL is a 
stochastic generalization of the deterministic method of lines approach\cite{bhmeadd} that relies on 
discretizing the spatial derivates without a temporal discretization, thus yielding to a set of ordinary 
differential equations in time, that can be easily integrated using the standard numerical 
libraries\cite{tspaper}. Though spatial accuracy can be increased by using spectral collocation 
method\cite{kabhadmen}, obtained solution is usually limited by the temporal accuracy of the integrator. 
SMOL is numerically stable without computational hindrance with convergence, is less computationally 
overloaded, spatiotemporally second order accurate, satisfies discrete FDT and can faithfully reproduce 
lab-based experiments {\it in silico}\cite{bhatnucl,chdutuyur,chtuyur,niskcorazumu,niskmus}. 

In coarsening kinetics, a kinetic length scale is extracted from the scattered light intensity 
inscribed within the structure function $S({\bf q}, t)$ as $\xi = [\int d^3{\bf q} S({\bf q},t) 
q^2/\int d^3{\bf q} S({\bf q},t)]^{-1/2}$, where 
$S({\bf q},t)={Q_{ij}({\bf q},t)Q_{ij}({\bf -q},t)/\int{d^3{\bf q}Q_{ij}({\bf q},t)Q_{ij}({\bf -q},t)}}$ 
with the orientation tensor  defined as ${\bf Q}({\bf q}, t) = \int_\mathbb{V}d^3x {\bf Q}({\bf x},t) 
e^{-i \bf q \cdot \bf x}/\mathbb{V}$ \cite{bray}. At equilibrium, the Ornstein-Zernicke form of the spatial 
correlation $\langle{\bf Q}(0){\bf Q}({\bf x})\rangle\propto\pmb\Gamma e^{-|{\bf x}|/\zeta}/|{\bf x}|$ 
defines the coherence length $\zeta = [\{1+2(\kappa+\Theta)/3\}L_1/A]^{1/2}$. By definition, $\zeta$ 
determines the core size of a disclination which is distinct from the length scale $\xi$ that denotes 
the mutual separation between two strings and also the separation between $\pm1/2$ integer dipoles 
within a shrinking disclination loop (or the radius of curvature). Thus, the length scale obtained 
from the disclination surface density $\rho\sim\xi^{-2}$ shown in frame \ref{fig:3}{\bf g} is identical 
with the length scale obtained from correlation functions\cite{bhatam}. In our numerics, time, length, 
and energy scales are resolved by non-dimensionalizing equations (\ref{eq:S1}-\ref{eq:S2})\cite{bhatnucl} 
and strictly retaining (i) {\it Courant-Friedrichs-Lewy} condition for timestep to avoid stiffness\cite{cfl}, 
(ii) $\Delta x\ll\zeta$ to resolve the grid spacing and distortion length embedded in the disclination 
core, (iii) ${\pmb\partial S<S/\zeta}$ for the validity of the GLdG method and, (iv) $k_BT\ll$ barrier 
height between isotropic-nematic state at clearing point to avoid any spurious oscillations between 
the two phases, thus, $k_BT = \mathcal{F}^{*}/10$, where $\mathcal{F}^{*} = 9C S_{ueq}^4/16$ is the 
non-dimensionalized distortion free energy\cite{bhatam}. With the chosen values of $\{\epsilon_0, 
\epsilon_s, \epsilon_a\}$, electric field relaxation is rapid compared to the orientational kinetics,
so as a steady electric field is obtained for each step of {\bf Q}-evolution. Grid size independence, 
numerical accuracy, and validity of physical tests were confirmed for each presented graphics and 
for both equations (\ref{eq:S1}-\ref{eq:S2}). 
\begin{figure}
\centering
\includegraphics[width=0.5\textwidth]{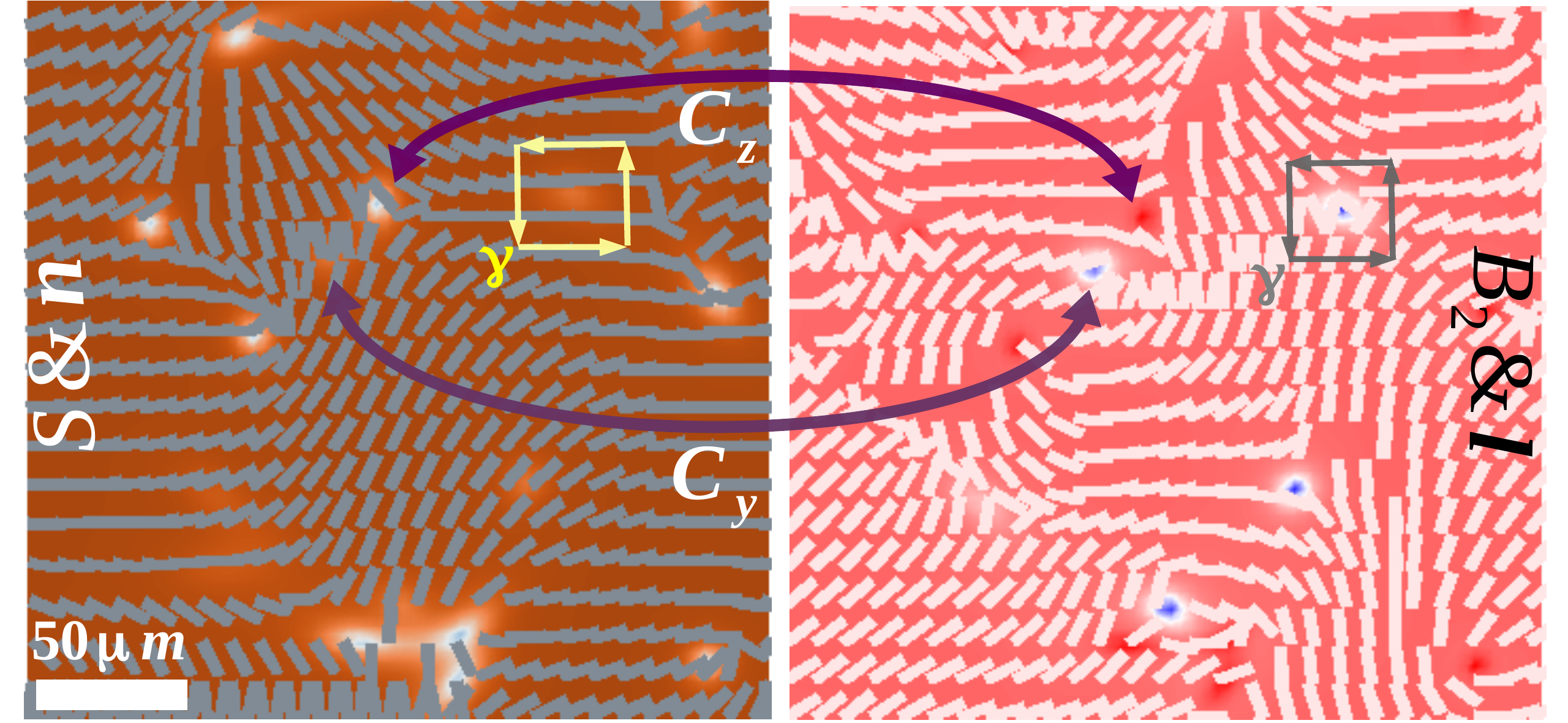}
\caption{\label{fig:S1} A portion of the ${\it xy}$-slice plane of an $80^2\times160{\mu}m^3$ thermotropic biaxial NLC (portrayed 
in frame \ref{fig:1}{\bf e,g}) displaying $\pm1/2$ integer defects corresponding to two different class 
of disclinations. Note that singular points of homotopy class $C_y$ are markedly different than $C_z$ 
by ${\pmb\partial \bf n}=0$, otherwise for both classes, $({\pmb\partial S,\pmb\partial B_2,\pmb\partial 
\bf l)}\neq0$. Material (computation) parameters are tabulated in Table \ref{tbl:GLdGparam}.}
\end{figure}

We finally discuss difference of our method with existing methodology in connection with the electrorheology 
of disclinations. Within Leslie-Ericksen (LE) theory in one dimension using free energy minimization 
technique, both the effect of rheology\cite{palepiboci} and nonuniform electric field\cite{cumecakon} 
has been exercised. Traditionally, LE theory is suitable in deep within the nematic phase where it is 
assumed that \{$S,B_2,{\bf l}$\} are constant in the orientation tensor and elastic distortion is concealed 
in ${\bf n}$ only. Earlier attempts to include hydrodynamics\cite{qishen} to LE theory were through the 
finite element method (FEM)\cite{jawifeday}. Also, the finite volume method (FVM) applied to Smoluchowski 
equation for the orientational probability density function is yet at a preliminary stage\cite{shrixipe}. 
On the other hand, after the advent of large scale computation, numerical solution of the athermal 
orientation tensor equation retaining all independent degrees of freedom is exercised through cell-dynamic 
scheme (CDS)\cite{zapgolgol}, finite difference methods (FDM) \cite{olmgol,aratan} as well as the method 
of lines (MOL)\cite{bhmeadd} approach to the GLdG theory. Attempts to include hydrodynamics and thermal 
fluctuations to the FDM via fluid particle dynamics (FPD)\cite{aratan} were replaced with the Lattice-Boltzmann method 
(LBM)\cite{limastca,denorlyeom1}, while inclusion of thermal fluctuations retaining second-order 
numerical accuracy is extended as a stochastic generalization of the method of lines (SMOL)\cite{bhmeads}. 
In this manuscript, SMOL complements the inclusion of Maxwell's equation to the existing formulation 
and motivates the association of hydrodynamics\cite{donobhgabe} for intense electric field studies 
and Fourier's law of latent heat conduction\cite{absorey} for confined NLC systems. \\

\section{\sc \bf Identification and topological classification of disclinations.}
\label{app:3}
Disclinations of different homotopy class are obtained after extracting \{$S,B_2,{\bf n},{\bf l}$\} from 
the basis coefficients $a_i$ on each space point by a similarity transformation. The scalar values are 
colour rendered according to the indicated bars in figure \ref{fig:1}. Disclinations are identified by 
sketching the isosurfaces with specific isovalue of the scalar fields. At a late stage of the kinetics 
when few isolated disclinations are existent, a plane from the three-dimensional volume is sliced in which 
the lateral section of disclinations appear as points. Once the physical location of disclinations are 
identified, the homotopy class, charge and sign of defect is calculated from the spatial distribution 
of \{{\bf n,l}\}. Similar to the Volterra construction in crystal dislocation, a Burgers circuit $\gamma$ 
displayed in figure \ref{fig:S1} is constructed using neighboring lattice points that encircle the 
defect\cite{zapgolgol}. The angular shift of \{{\bf n,l}\} is measured while traversing one complete 
loop and the charge and sign of defects are estimated using the hodograph method\cite{klelav}. If 
${\bf u}=({sin}\theta {cos}\phi,{sin}\theta {sin}\phi,{cos}\theta)$ 
is directionally equivalent with ${-\bf u}=({sin}\theta^\prime {cos}\phi^\prime,
{sin}\theta^\prime {sin}\phi^\prime,{cos}\theta^\prime)$ where $(\theta,\phi)$ 
are polar and azimuthal angle in an arbitrary frame, then following transformation retains the 
centrosymmetry,
\begin{align}
\label{eq:S3}
\theta^\prime \to \pi - \theta; \; \phi^\prime \to \pi + \phi. 
\end{align}
For $\pm1/2$ integer disclinations in a slice plane of uniaxial NLC shown in frame \ref{fig:1}{\bf c}, 
while traversing $\gamma$, {\bf n} rotates by $\pm\pi$. In biaxial NLC, homotopy classes are identified 
using the following recipe\cite{zapgolgol}: (i) for $C_x$ class of disclinations, ${\bf n}$ rotates by 
$\pm\pi$ but ${\bf l}$ does not rotate, (ii) for $C_y$ class of disclinations, ${\bf n}$ does not rotate 
but ${\bf l}$ rotates by $\pm\pi$ and, (iii) for $C_z$ class of disclinations, both $\{\bf n,l\}$ rotates 
by $\pm\pi$. We do not find any $C_x$ class of disclinations, which is consistent with the analytic 
prediction on two-dimensional nonabelian vortices\cite{kobtho} and numerical computations\cite{zapgolgol,
bhatam}. This algorithm not only supersedes the traditional defect classifying approaches using vector 
field, tensor glyph or hyperstreamline seeding through Mueller and Westin matrices\cite{fuabukh,calpellor} 
but can uniquely determine disclinations from the structure of orientation tensor rather than rely on 
the vectorial information.


\newpage
\widetext
\begin{center}
\textbf{\large Supplemental Information}
\end{center}

\section*{Procedure to sample the field variables}
As an initial condition, we prepare the isotropic state by drawing $S$ and $T$ randomly from a normal 
distribution $\mathcal{N}(0, 2k_BT)$ using Box–Muller transform to the standard library routines\cite{tspaper}. 
We sample $cos\theta$ from a uniform distribution between $-1$ and $1$, $\phi$ between $0$ and $2\pi$ 
to generate director ${\bf n}$ and codirector ${\bf l}$. The secondary director ${\bf m}$ is constructed 
through the Gram-Schmidt orthogonalization procedure\cite{arfweb}. The thermal fluctuations are incorporated 
in each numerical step using the basis transformation ${\pmb \xi} \rightarrow \zeta$\cite{bhmeads}. After 
making the transformation ${\bf Q} \rightarrow a$, we evolve equation(\ref{eq:S1}), presented in the Appendix 
section in the main article, for different initial configurations and switch on and off the electric potential 
difference at $t_{on}$ and $t_{off}$ to sample the nonequilibrium state space, as the system transforms 
from an isotropic phase to a nematic phase. Using an inverse transformation $a \rightarrow {\bf Q}$ and 
space derivate of the electric potential $\Psi$, we reconstruct the orientation tensor and electric flux 
lines at every space point. 

\section*{Comparison between uniform and nonuniform electric field scenario}
\renewcommand{\figurename}{Supplementary Figure}
\renewcommand{\thefigure}{S\arabic{figure}}
\setcounter{figure}{0} 
\begin{figure}
\centering
\includegraphics[width=0.5\textwidth, height=0.45\textwidth]{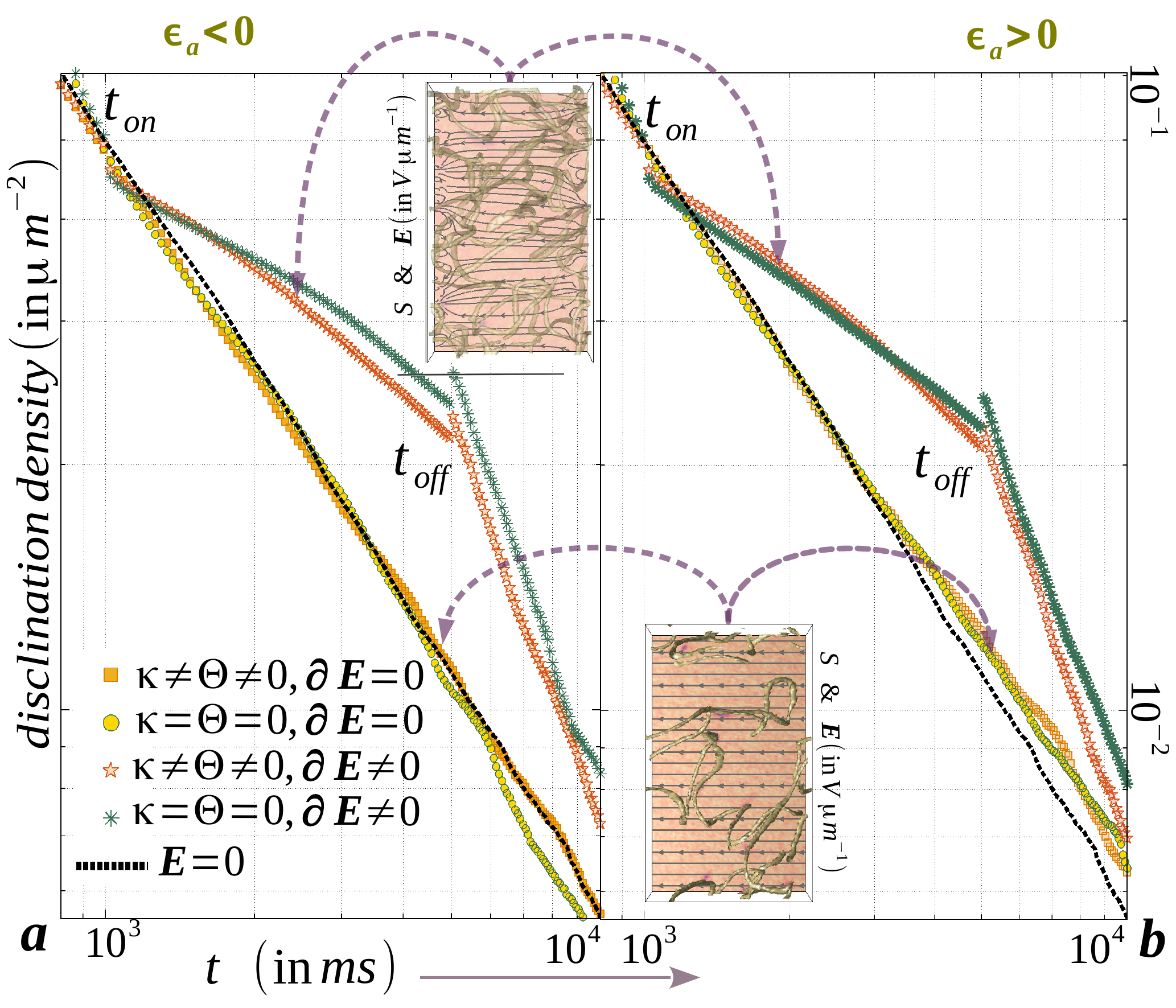}
\caption{\label{fig:F1} Comparison of the kinetics of disclination density per unit area for uniform 
[solution of equation (7)] and nonuniform [solution of equation (6)] electric field, together with the 
degenerate (one) elastic constant approximation ($\kappa=\Theta=0$) and non-degenerate variant 
($\kappa\neq\Theta\neq0$) in a coarsening thermal uniaxial NLC media. Panel {\bf a} (Panel {\bf b}) 
display the same for negative (positive) dielectric constant material. Though the kinetics for degenerate 
and non-degenerate elastic constant scenario are very similar, an increased life-span of $\pi$-solitons 
with non-Markovian response during field cessation is observed when nonuniformity in the electric field 
is considered.}
\end{figure}

In Supplementary Figure \ref{fig:F1}, we compare the decay of surface disclination density with time 
for thermal uniaxial media for both sign of dielectric constant for various cases: (i) equilibrium 
disclination kinetics (electric field {\bf E} = 0), disclination network under (ii) uniform and (iii) 
nonuniform {\bf E} for equal ($\kappa=\Theta=0$) and unequal ($\kappa\neq\Theta\neq0$) elastic constant 
approximation. Field switch on ($t_{on}$) and switch off ($t_{off}$) with an instantaneous snapshot of the 
uniaxial degree along with the electric flux lines are also embedded in the graphics. Similar to figure 
2{\bf c} in the main article, we find that far from the critical line, the response of the disclinations 
to the electric field for degenerate and non-degenerate elastic constants are similar to statistical 
averaging. However, in the presence of nonuniform electric field for both degenerate and non-degenerate
elastic constant scenario, the disclinations are long-lived. 

\bibliographystyle{apsrev}
\bibliography{references}
\renewcommand{\figurename}{Supplementary Movie}
\renewcommand{\thefigure}{S\arabic{figure}}
\setcounter{figure}{0} 
\begin{figure*}[!b]
\centering
\includegraphics[width=0.45\textwidth, height=0.3\textwidth]{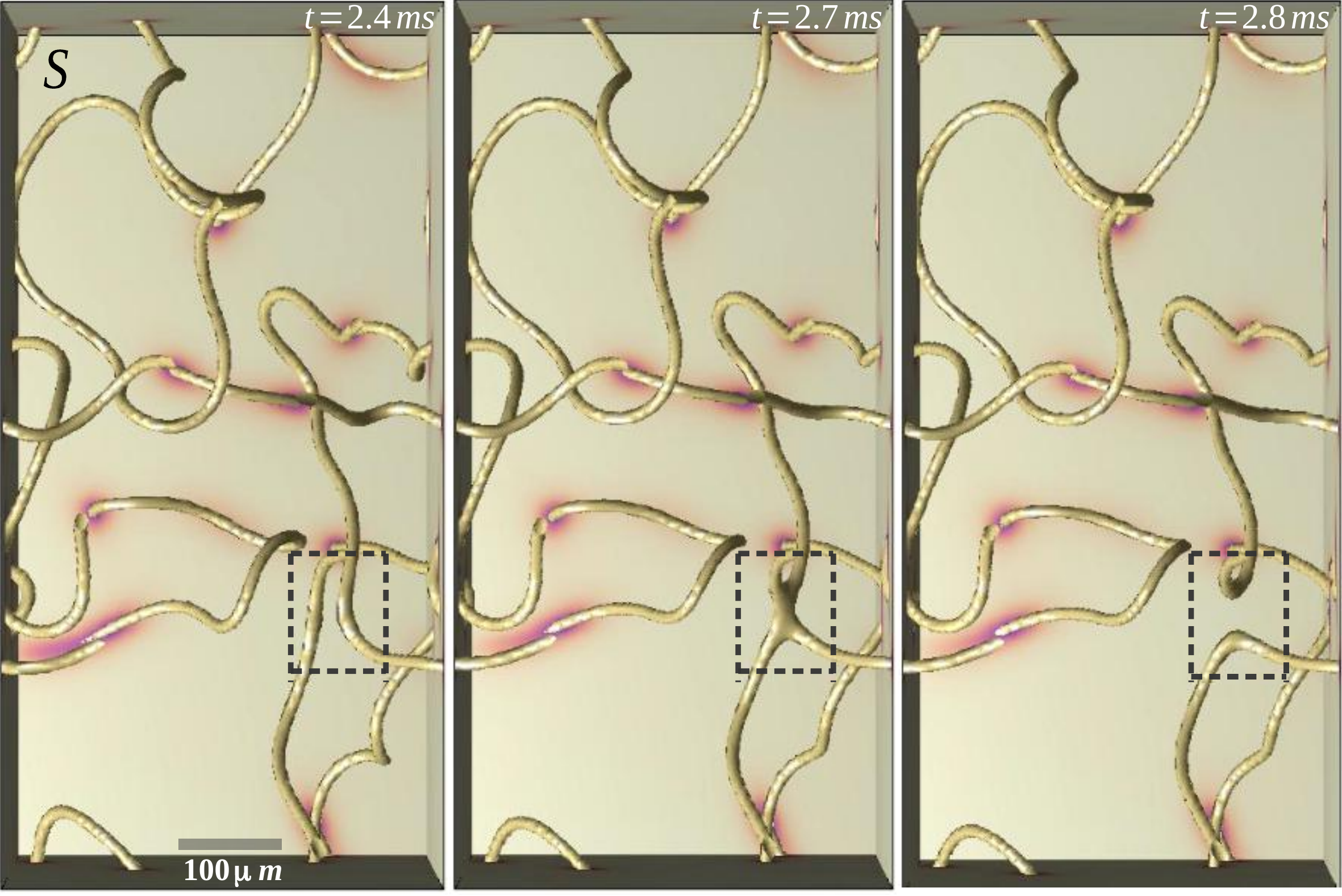}
\caption{\label{fig:M1} Nucleation, intercommutation (marked with dashed box) and extinction of disclinations 
in a coarsening athermal uniaxial NLC. The animation sequentially portrays of the evolution of uniaxial order 
$S$ \& biaxial order $B_2$ for an uniaxial NLC subjected to a temperature quench from a disordered isotropic 
phase to an ordered uniaxial nematic phase. Colourbars and isovalues corresponding to the isosurfaces of 
disclinations are indicated in figure $1$ in the main article. As $S$ grows towards the saturation value 
$S_{ueq}$, isotropic domain coarsening leads to the nucleation of charge neutral disclinations with $\pm1/2$ 
integer topological dipoles at two end segments at around $1.2ms$. These strings intercommute by exchanging 
segments with each other to form closed contractile loops that are extinct from the medium to minimize the 
total (free) energy. The right panel displays of the reduced $B_2$ in the media with $(B_2)_{max}$ around 
the disclination core.}
\end{figure*}

\begin{figure*}
\centering
\includegraphics[width=0.75\textwidth, height=0.2\textwidth]{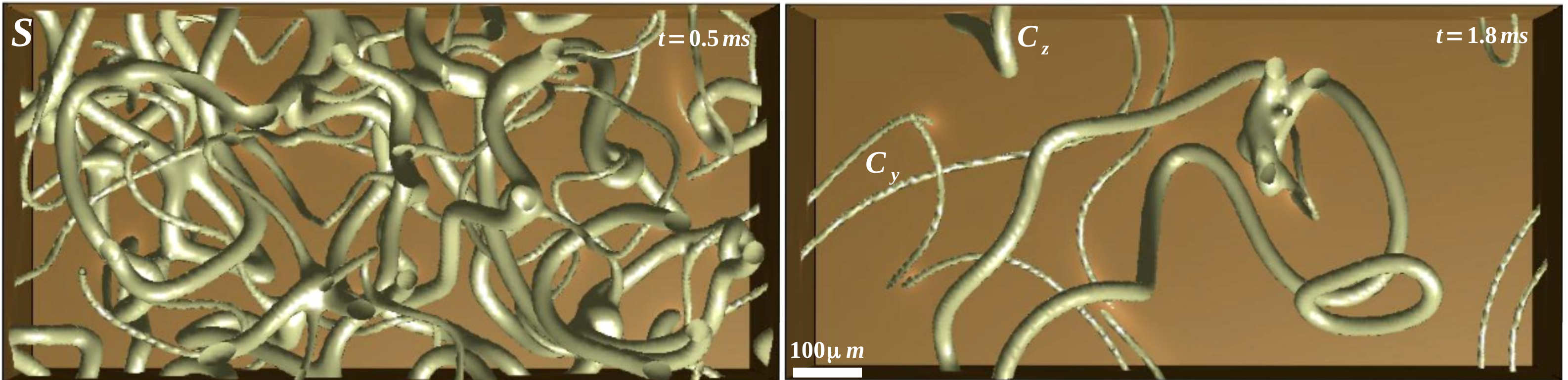}
\caption{\label{fig:M2} Disclination kinetics in a coarsening athermal biaxial NLC. Similar to Supplementary 
Movie \ref{fig:M1}, this animation displays of the evolution of $S$ \& $B_2$ for a thermotropic biaxial NLC, 
subjected to a temperature quench from an isotropic phase deep into the biaxial phase. Colourbars and isovalues 
corresponding to the isosurfaces of disclinations are indicated in figure $1$ in the main article. As \{$S,B_2$\} 
grows in the media towards the saturation value \{$S_{beq},(B_2)_{beq}$\}, isotropic domain coarsening leads 
to the nucleation of $\pi$ solitons of $C_y$ and $C_z$ class at around $0.5ms$ (left panel), that intercommute 
only within the respective class to form contractile loops to squeeze (right panel). In $B_2$, strings of similar 
width for different isovalues as mentioned in figure $1$ is coloured to distinguish. As already noted that 
devoid of chirality, the nonabelian disclinations of different class do not entangle but pass through each 
other.}
\end{figure*}

\begin{figure*}
\centering
\includegraphics[width=0.65\textwidth, height=0.3\textwidth]{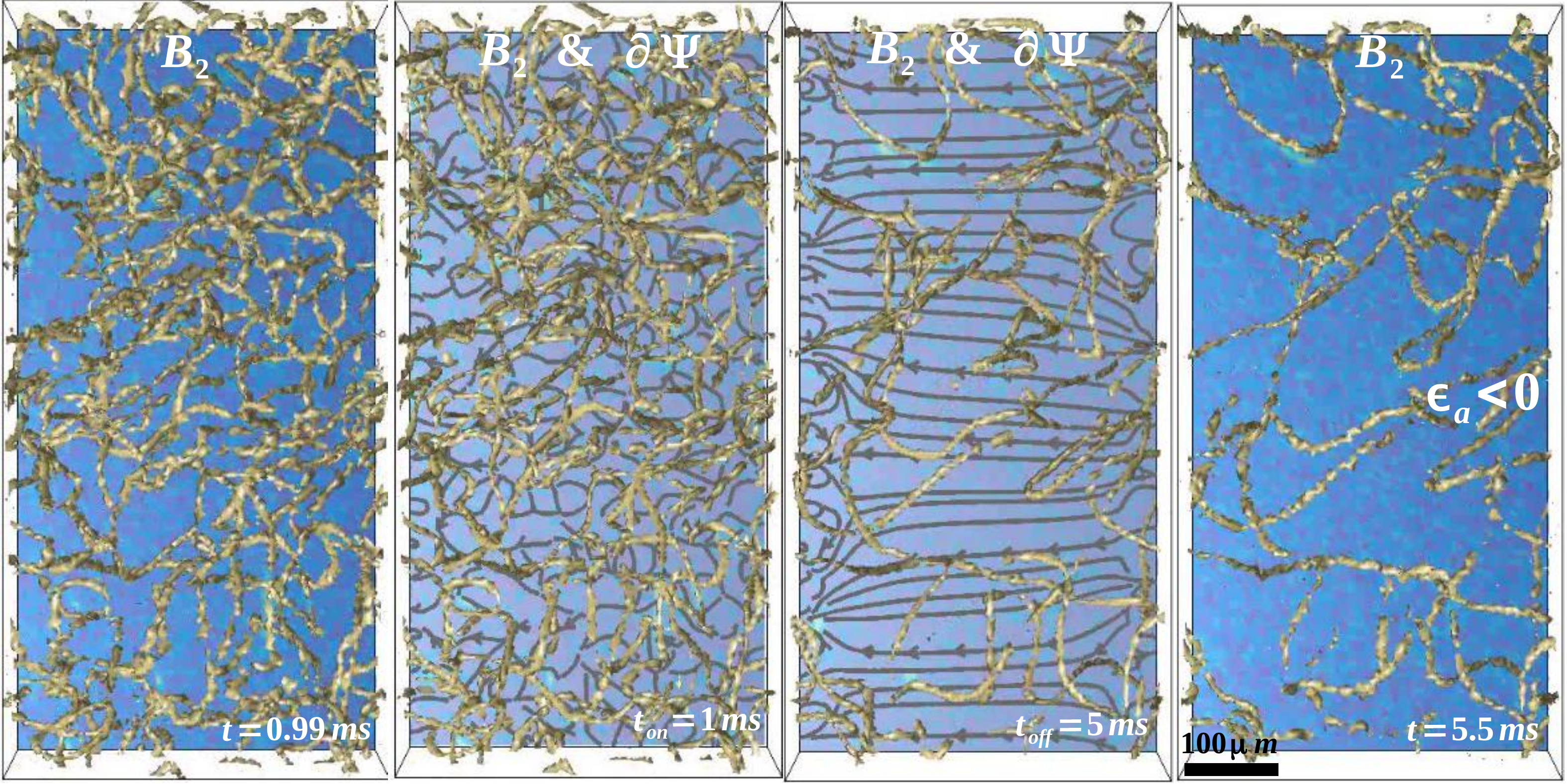}
\caption{\label{fig:M3} Nonuniform electric flux lines and temporal dilation of disclination kinetics in a 
coarsening thermal uniaxial NLC. The animation sketches the evolution of ($S,{\pmb\partial}\Psi$) \& 
($B_2,{\pmb\partial}\Psi$) for a thermal uniaxial NLC coarsening following a temperature quench from an isotropic 
state, subjected to the onset of an electric field at $t_{on}=1ms$ and cessation at $t_{off}=5ms$ for $\epsilon_a<0$. 
Similar response for $\epsilon_a>0$ is not shown for brevity. Note that in the second panel after $t_{on}$, electric 
field drives the fluctuating nematic media to attain a nematic phase with $S>S_{ueq}$, however, the line defect 
kinetics is significantly reduced during the interval $[t_{on},t_{off}]$. Note that disclinations are thinner 
after application of field and regains its equilibrium thickness after cessation of the field.}
\end{figure*}

\begin{figure*}
\centering
\includegraphics[width=0.65\textwidth, height=0.3\textwidth]{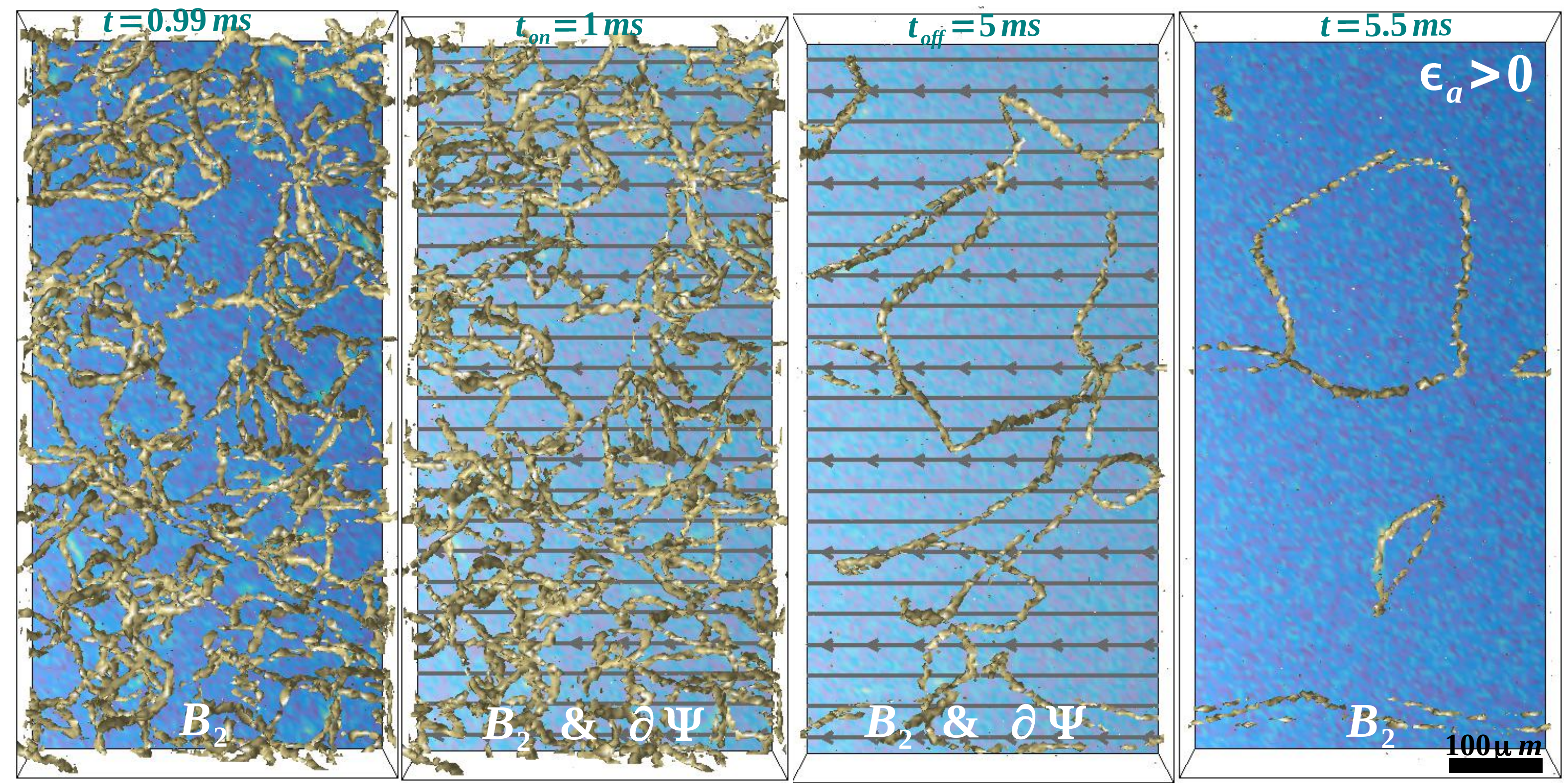}
\caption{\label{fig:M3a} Application of a uniform electric field, leading to deficient slowing down of the disclination 
kinetics in a coarsening thermal uniaxial NLC. The animation sketches the evolution of ($S,{\pmb\partial}\Psi$) \& 
($B_2,{\pmb\partial}\Psi$) for a thermal uniaxial NLC coarsening following a temperature quench from an isotropic 
state, subjected to the onset of an electric field at $t_{on}=1ms$ and cessation at $t_{off}=5ms$ for $\epsilon_a>0$. 
Similar response for $\epsilon_a<0$ is not shown for brevity. When compared to Supplementary Movie \ref{fig:M3}, 
the temporal dilation of disclination network in the presence of electric field is not substantial.}
\end{figure*}

\begin{figure*}
\centering
\includegraphics[width=1.0\textwidth, height=0.32\textwidth]{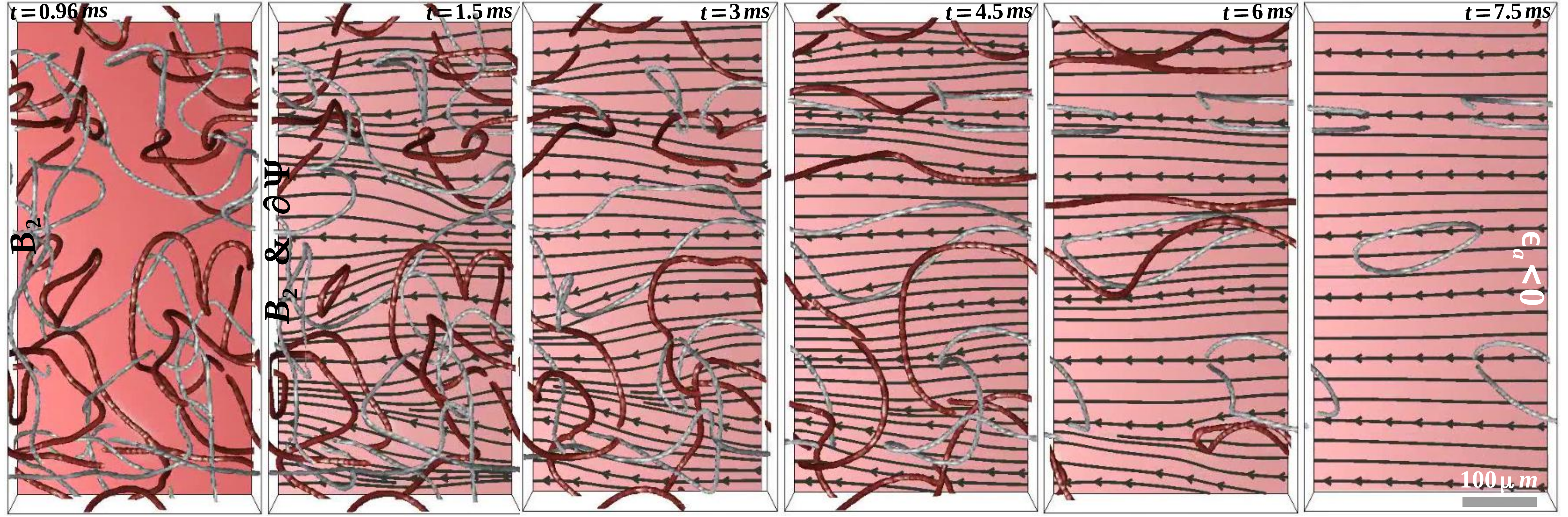}
\caption{\label{fig:M4} Nonuniform electric flux lines, temporal dilation of disclination kinetics and selection 
of $C_z$ class in a coarsening thermal biaxial NLC with {$\bf\pmb \epsilon_a<0$}. The animation sequentially 
portrays evolution of ($S,{\pmb\partial}\Psi$) \& ($B_2,{\pmb\partial}\Psi$) for a coarsening thermal biaxial 
NLC following a quench from an isotropic state, subjected to the onset of an electric force at $t_{on}=1ms$. 
Note that unlike Supplementary Movie S3, we do not switch off the field ($t_{off}\to\infty$). After $t_{on}$, 
the electric field drives the isotropic media to attain a biaxial nematic phase with $S>S_{beq}$, however, the 
disclination kinetics is sufficiently reduced. The disclinations are thinner after application of electric field. 
The kinetic pathway of $C_{y,z}$ become asymmetric and as a result, the $C_y$ class is long-lived. Selection of 
$C_y$ class is also seen in the solution of equation $6$ in the main article (not shown). We have also noted that 
the equilibrium dynamics is regained after cessation of field (not shown).}
\end{figure*}

\begin{figure*}
\centering
\includegraphics[width=1.0\textwidth, height=0.32\textwidth]{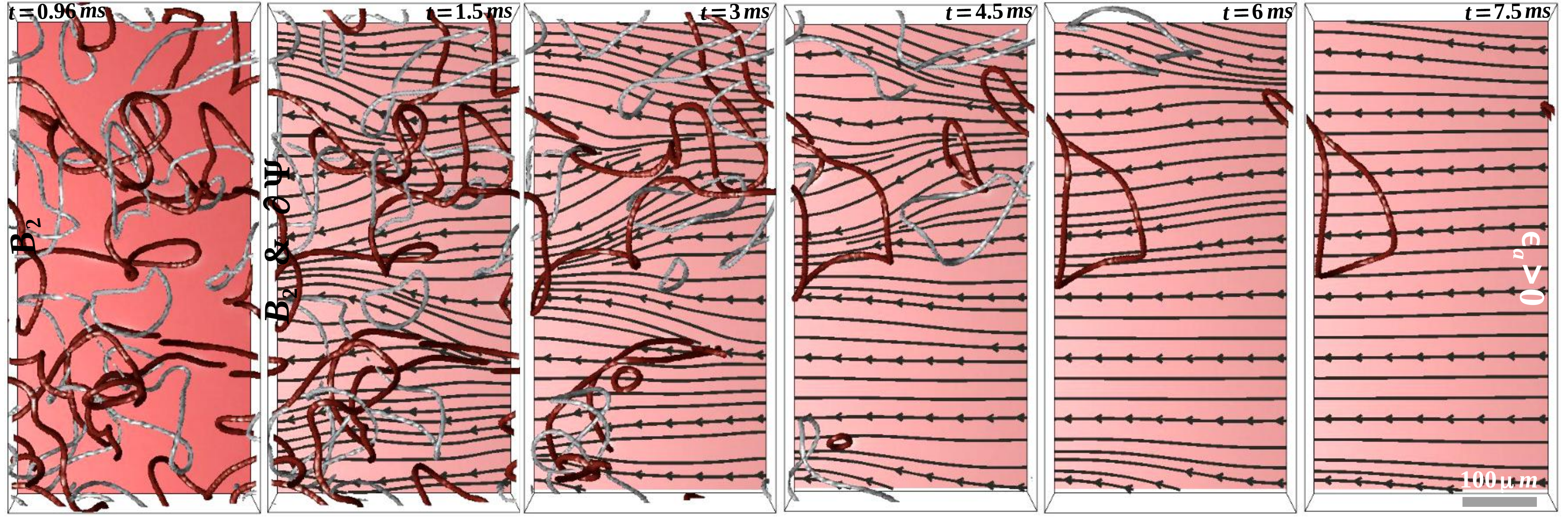}
\caption{\label{fig:M5} Nonuniform electric flux lines, temporal dilation of disclination kinetics and selection 
of $C_y$ class in a coarsening thermal biaxial NLC with {$\bf\pmb \epsilon_a>0$}. Instead of Supplementary Movie 
\ref{fig:M4}, here we find that $C_y$ class is long lived.}
\end{figure*}
\end{document}